\documentclass[a4paper,12pt]{article}
\usepackage{amsfonts,amsthm,amsmath,amssymb,graphicx}
\usepackage{graphicx,color}
\usepackage{hyperref}
\usepackage{enumitem}
\usepackage{mathrsfs,stmaryrd,mixing}
\usepackage{multirow,diagbox}

\numberwithin{equation}{section}

\let\OLDthebibliography\thebibliography
\renewcommand\thebibliography[1]{
  \OLDthebibliography{#1}
  \setlength{\parskip}{2pt}
  \setlength{\itemsep}{1.7pt plus 1.7pt}
}

\begin{document}
\thispagestyle{empty}

\ \vskip 50mm

{\LARGE 
\centerline{\bf Refined Counting of Necklaces}

\vspace{3mm}
\centerline{\bf in One-loop $\cN=4$ SYM}

}

\vskip 20mm

\centerline{
{\large \bf Ryo Suzuki}
}

{\let\thefootnote\relax\footnotetext{{\tt rsuzuki.mp\_at\_gmail.com}}}

\vskip 20mm

\centerline{{\it ICTP South American Institute for Fundamental Research},}
\centerline{{\it Instituto de F\'isica Te\'orica, UNESP - Universidade Estadual Paulista},}
\centerline{{\it Rua Dr. Bento Teobaldo Ferraz 271, 01140-070, S\~ao Paulo, SP, Brazil}}

\vskip 20mm


\centerline{\bf ABSTRACT}

\vskip 6mm 

We compute the grand partition function of $\cN=4$ SYM at one-loop in the $SU(2)$ sector with general chemical potentials, extending the results of P\'olya's theorem.
We make use of finite group theory, applicable to all orders of perturbative $1/N_c$ expansion. 
We show that only the planar terms contribute to the grand partition function, which is therefore equal to the grand partition function of an ensemble of XXX$_\frac12$ spin chains.
We discuss how Hagedorn temperature changes on the complex plane of chemical potentials.

\newpage

\tableofcontents

\setcounter{page}{0}
\setcounter{tocdepth}{2}
\setcounter{footnote}{0}

\newpage

\section{Introduction}\label{sec:intro}

The $\cN=4$ super Yang-Mills theory (SYM) has attracted a lot of attention owing to its simple and profound structure. Besides being the primary example of the AdS/CFT correspondence \cite{Maldacena:1997re}, this theory is believed to be integrable in the planar limit \cite{Minahan:2002ve}. The integrability enables us to predict various observables at any values of the 't Hooft coupling; see \cite{Beisert:2010jr} for a review.

As a parallel development, alternative methods have been developed to uncover the non-planar structure of $\cN=4$ SYM with the gauge group $U(N_c)$ or $SU(N_c)$, based on finite group theory \cite{Corley:2001zk}. New bases of gauge-invariant operators have been discovered, which diagonalize the tree-level two-point functions at finite $N_c$ \cite{Brown:2007xh,Bhattacharyya:2008rb,Bhattacharyya:2008xy,
Brown:2008ij}; see \cite{Ramgoolam:2016ciq} for a review.

With numerous approaches at hand to study individual operators, let us ask questions complementary to the above line of development. We reconsider the statistical property of $\cN=4$ SYM, namely the grand partition function including perturbative $1/N_c$ corrections.

In \cite{Sundborg:1999ue}, the tree-level partition function of $\cN=4$ SYM on $\bb{R} \times {\rm S}^3$ was computed to investigate its phase space structure. 
The free energy has the expansion $F = N_c^2 \, F_0 + F_1 + \dots$\,, where $F_0=0$ in the confined or low-temperature phase, and $F_0>0$ in the deconfined or high-temperature phase. In the confined phase, the density of states increases exponentially as the energy increases, leading to the singularity of the partition function at a finite temperature. 
There the vacuum undergoes the so-called Hagedorn transition to the deconfined phase \cite{Aharony:2003sx,Yamada:2006rx}.
In the dual supergravity, it is argued that a thermal scalar in \AdSxS\ becomes tachyonic at a finite temperature, and condensates into the AdS blackhole \cite{Kruczenski:2005pj}.

Below we consider the grand partition function of $\cN=4$ SYM in the low-temperature phase at one-loop. It amounts to summing up the one-loop anomalous dimensions of all gauge-invariant operators. The problem simplifies a lot by noticing that we do not need to take an eigenbasis of the dilatation operator to compute the trace. The main problem is how to take the trace efficiently in the general setup.
The one-loop partition function without chemical potential has been obtained by using P\'{o}lya enumeration theorem in \cite{Spradlin:2004pp}. 
The Hagedorn transition in the pp-wave/BMN limit was studied in \cite{PandoZayas:2002hh,Grignani:2003cs}.
The grand partition function with a one-parameter family of chemical potential was given in \cite{Harmark:2006di}, and the phase space near the critical chemical potentials was studied in \cite{Harmark:2014mpa}.

However, at one-loop the P\'olya-type formulae are known only for single-variable cases, which makes it difficult to obtain the grand partition function with general chemical potentials. In this paper, we incorporate the fully general chemical potentials in the $SU(2)$ sector, by using finite group theory which is valid to all orders of perturbative $1/N_c$ expansion.\footnote{A similar quantity was computed in \cite{Sochichiu:2006uz,Sochichiu:2006yv}, that is the grand partition function with general chemical potential at one-loop in the $SU(2)$ sector including $O(N_c^2)$ term. This result involves a multi-dimensional integral, and less explicit than our counting formula. Another formula was obtained in \cite{Bianchi:2006ti}, namely the trace of the one-loop dimension over the product of two fundamental representations of $\alg{psu}(2,2|4)$. This quantity is a building block in the P\'olya-type formula, but not identical to the grand partition function.}

In the planar limit, the dilatation operator of $\cN=4$ SYM at one-loop in the $SU(2)$ sector is the Hamiltonian of XXX$_{\frac12}$ spin chain. The (canonical) partition function of the spin chain can be computed by using string hypothesis in the thermodynamic limit. Our work provides alternative derivation based on the microscopic counting of states, if we take care of subtle differences to be discussed in Section \ref{sec:BAE}.

\paragraph{Main results.}

Let us briefly summarize the main results of this paper. We consider the grand partition function of $\cN=4$ SYM in the confined phase, with the gauge group $U(N_c)$ up to one-loop in the $SU(2)$ sector.
The theory is put on $\bb{R} \times {\rm S}^3$\,, where $\bb{R}$ is the radial direction of $\bb{R}^4$ and the radius of ${\rm S}^3$ is set to unity.
We write the grand partition function as
\begin{equation}
\cZ (\beta, \vec \omega) = \tr \( e^{-\beta \fD + \sum_i \omega_i J_i} \), \qquad
\fD = \fD_0 + \lambda \fD_2 + \dots ,
\label{def:spt part fn}
\end{equation}
where $\fD$ is the dilatation operator and $J_i$ are the R-charges.
We take the trace over all gauge-invariant local operators in the $SU(2)$ sector, made out of complex scalars $\{ W,Z \}$.
The grand partition function \eqref{def:spt part fn} has the weak coupling expansion
\begin{equation}
\cZ (\beta, x,y) = Z_0^{\rm MT} (x, y) - 2 \lambda \beta Z_2^{\rm MT} (x, y) + O(\lambda^2), \qquad
\lambda = \frac{N_c \, g_{\rm YM}^2}{16 \pi^2} \,,
\label{def:expand cZxy}
\end{equation}
where the operators $W^m Z^n$ are weighted by $x^m y^n$.
We apply finite group theory to compute (the perturbative part of) $Z_0^{\rm MT}$ and $Z_2^{\rm MT}$.
It will turn out that only the planar term contributes at one-loop in this setup, even though our methods are valid to all orders of perturbative $1/N_c$ corrections.

On top of $1/N_c$ corrections, there are non-perturbative corrections coming from finite $N_c$ constraints. At finite $N_c$\,, certain combinations of operators vanishes when their canonical dimension of an operator exceeds $N_c$\,. 
Let us write the exact grand partition function as
\begin{equation}
\cZ^{\rm exact}_{N_c} (\beta, x,y) = \sum_{m,n \ge 0} \mathscr{N}_{m,n}^{(p)} (N_c) \, x^m y^n -
\sum_{\substack{m,n \ge 0 \\m+n > N_c}} \mathscr{N}_{m,n}^{(np)} (N_c) \, x^m y^n ,
\label{cZ finite Nc}
\end{equation}
where the second sum represents the subtraction at finite $N_c$\,. If we set $x=y=e^{-\beta}$, then the second sum is of order $e^{-\beta N_c}$ at large $\beta$\,, which are non-perturbative in view of $1/N_c$ expansion. Our methods capture the first sum in \eqref{cZ finite Nc}, which may contain the corrections of order $1/N_c^\ell$ at $\ell$-loop.
To simplify the notation, we write e.g. $Z_0^{\rm MT} (x, y)$ instead of $Z_0^{{\rm MT} (p)} (x, y)$ throughout the paper.

\bigskip
We obtain two expressions of $Z_2^{\rm MT}$ in Section \ref{sec:one-loop count}, which we call Partition form and Totient form. The $Z_2^{\rm MT}$ in Partition form is written as a sum over partitions of operator length $L$\,, $r=[1^{r_1}, 2^{r_2}, \dots, L^{r_L}]$\,:
\begin{multline}
Z_2^{\rm MT} (x, y) = N_c \sum_{L=0}^\infty \sum_{r \vdash L} \,
\prod_{k=1}^\infty (x^k + y^k)^{r_k} \ \Big\{ L - \sum_{a=1}^L \theta_> (r_a)
- \sum_{a=1}^{L/2} a \, (r_a+1) \theta_> (r_{2a}) 
\\
- 2 \sum_{a< b}^L \theta_> (L+1-a-b) \theta_> (r_a) \theta_> (r_b)  
- \sum_{a=1}^{L/2} \theta_> (r_a-1) 
\Big\},
\label{main Z2mt partition}
\end{multline}
where $\theta_>(x)=1$ for $x > 1$, and vanishes otherwise. The $Z_2^{\rm MT}$ in Totient form involves Euler's totient function ${\rm Tot} (d)$, which counts the number of relatively prime positive integers less than $d$\,:
\begin{multline}
Z_2^{\rm MT} (x, y) = N_c \, \prod_{h=1}^\infty \frac{1}{1- x^h - y^h} \ \times 
\\
\Bigg[
\sum_{k=1}^\infty 
2 \( \sum_{d=1}^\infty {\rm Tot} (d) \frac{x^{kd} y^{kd}}{1- x^{kd} - y^{kd}}
- \sum_{L=2}^\infty \sum_{m=1}^{L-1} x^{km} y^{k(L-m)} \, \delta({\rm gcd}(m,L), 1) \)
\Bigg] .
\label{main Z2mt totient}
\end{multline}
The equivalence of two results can be checked by expanding both series around the zero temperature, corresponding to the limit of small $x,y$.
When the gauge group is $SU(N_c)$, an overall factor of $(1-x)(1-y)$ should be multiplied.

It is straightforward to compute the Hagedorn temperature by using Totient form \eqref{main Z2mt totient}. 
The Hagedorn temperature $T_H(\lambda)$ in the $SU(2)$ sector is a function of the chemical potentials $(\omega_1 , \omega_2) = (\log x + \frac{1}{T}, \, \log y + \frac{1}{T})$, which is given by
\begin{equation}
T_H (\lambda) = 
\begin{cases}
\ds \frac{1}{{\log (e^{\omega_1}+e^{\omega_2})}} \[ 1 + 
\frac{4 \lambda  e^{\omega_1 + \omega_2}}{\left(e^{\omega_1}+e^{\omega_2}\right)^2} \] 
&\quad \( e^{\omega_1} > 0 \ \ {\rm and} \ \ e^{\omega_2} >0 , \ e^{\omega_1}+e^{\omega_2} \ge 1 \),
\\[5mm]
\ds \frac{2}{\log (e^{2 \omega_1}+e^{2 \omega_2} )} \[ 1 + 
\frac{4 \lambda  e^{2 \omega_1 + 2 \omega_2}}{(e^{2 \omega_1}+e^{2 \omega_2})^2}
\]   
&\quad \( e^{\omega_1} < 0 \ \ {\rm or} \ \  e^{\omega_2} < 0 , \ e^{2\omega_1}+e^{2\omega_2} \ge 1 \),
\end{cases}
\label{main TH}
\end{equation}
which is valid at large $N_c$\,, due to the second sum in \eqref{cZ finite Nc}.
Roughly said, the first line represents the deconfinement of $W$ and $Z$, whereas the second line that of $W^2$ and $Z^2$. This result shows that the complex chemical potentials change the location of the Hagedorn transition, as will be discussed further in Section \ref{sec:Hagedorn}.

\section{Tree-level counting}\label{sec:tree count}

We introduce two methods of computing the generating function of the number of gauge-invariant operators in $\cN=4$ SYM with $U(N_c)$ gauge group. This generating function is equal to the grand partition function at tree-level.

\subsection{Permutation basis of gauge-invariant operators}\label{sec:pm basis}

General gauge-invariant local operators of $\cN=4$ SYM can be specified by an element of permutation group.
We introduce the elementary fields of $\cN=4$ SYM
\begin{equation}
\cW^A=\{ \nabla^s \Phi^I ,\nabla^s F, \nabla^s \bar{F}, \nabla^s \psi ,\nabla^s \bar{\psi} \}, \qquad (s \ge 0),
\end{equation}
and their polynomials
\begin{equation}
\begin{aligned}
\cO_\alpha^{A_1\dots A_L} 
&= \tr_L \[ \alpha \, \cW^{A_1} \cW^{A_2} \dots \cW^{A_L} \] ,
\\
&= \sum_{a_1, a_2, \dots, a_L=1}^{N_c}
(\cW^{A_1})^{a_1}_{a_{\alpha(1)}} (\cW^{A_2})^{a_2}_{a_{\alpha(2)}} \dots (\cW^{A_L})^{a_L}_{ a_{\alpha(L)}} \,, \qquad
(\alpha \in S_L).
\end{aligned}
\label{def:generic multi-trace}
\end{equation}
There is an equivalence relation coming from the relabeling $(a_i \,, A_i) \to (a_{\gamma(i)} \,, A_{\gamma(i)})$,
\begin{equation}
\cO_\alpha^{A_1\dots A_L} =
\cO_{\gamma \alpha \gamma^{-1}}^{ A_{\gamma(1)},\dots, A_{\gamma(L)} } , \qquad
(\forall \gamma \in S_L).
\label{def:equiv relation}
\end{equation}
Each of the equivalence class uniquely specifies a gauge-invariant operator.

In the $SU(2)$ sector, we restrict $\cW^A$ to the elementary fields to a pair of complex scalars $\{W, Z\}$. We use the gauge degrees of freedom \eqref{def:equiv relation} to set
\begin{equation}
\cO_\alpha = \tr_{L} ( \alpha \, W^m Z^n ), \qquad
\cW^{A_i} = \begin{cases}
W &\quad (1 \le i \le m) \\
Z &\quad (m+1 \le i \le m+n) .
\end{cases}
\label{def:su2op mn}
\end{equation}
The residual gauge degrees of freedom is
\begin{equation}
\cO_{\alpha} = \cO_{\gamma \alpha \gamma^{-1}} \,, \qquad
(\forall \gamma \in S_m \times S_n ).
\label{su2 symm SmSn}
\end{equation}
This equivalence class uniquely specifies a gauge-invariant operator in the $SU(2)$ sector.
The number of such multi-trace operators at a fixed $(m,n)$ is given by summing over all solutions of \eqref{su2 symm SmSn},\footnote{This formula is called Burnside's lemma. The number of operators is related to the large $N_c$ limit of the tree-level two-point functions \cite{Kimura:2016bzo}.}
\begin{equation}
N^{\rm MT}_{m,n} = \frac{1}{m! \, n!} \sum_{\alpha \in S_{m+n}} \sum_{\gamma \in S_m \times S_n} \delta_{m+n} (\alpha^{-1} \gamma \alpha \gamma^{-1}) ,
\label{def:NSU2 mn}
\end{equation}
where
\begin{equation}
\delta_L (\sigma) = \begin{cases}
1 &\qquad (\sigma = 1 \in S_L) \\
0 &\qquad ({\rm otherwise})
\end{cases}
\end{equation}

The generating function of the number of multi-trace operators is defined by
\begin{equation}
Z_0^{\rm MT}(x,y) = \sum_{m,n=0}^\infty N_{m,n} \, x^m y^n ,
\label{def:Z0mt}
\end{equation}
which will be computed below.

\subsection{Sum over partitions}\label{sec:tree partition}

First we evaluate $N^{\rm MT}_{m,n}$ in \eqref{def:NSU2 mn}.
Let us write $\gamma = \gamma_W \cdot \gamma_Z$ with $\gamma_W \in S_m$ and $\gamma_Z \in S_n$\,. Suppose that $\gamma_W \,, \gamma_Z$ have the cycle structure $p \vdash m, q \vdash n$, respectively.
\begin{equation}
p \vdash m \qquad \Leftrightarrow \qquad
p = [1^{p_1}, 2^{p_2} , \dots m^{p_m}], \qquad
\sum_{k=1}^m k \, p_k  = m.
\label{def:partition exponential}
\end{equation}
We also define
\begin{equation}
r_k = p_k + q_k \,, \qquad r \vdash m+n.
\label{def:rk}
\end{equation}

We look for the general solutions of the condition $\alpha^{-1} \gamma \alpha \gamma^{-1} = 1$ for a fixed $\gamma$, which we call stabilizer ${\rm Stab} (\gamma)$. Let us parametrize $\gamma$ by
\begin{equation}
\gamma = \prod_{k=1}^{m+n} \prod_{h=1}^{r_k} (g_{h,1}^{(k)} \dots g_{h,k}^{(k)}),
\label{parametrize gamma}
\end{equation}
where $(g_{h,1}^{(k)} \dots g_{h,k}^{(k)})$ is the cyclic permutation defined in Appendix \ref{app:notation}. The identity \eqref{perm identities} gives
\begin{equation}
\alpha^{-1} \gamma \alpha = \prod_{k=1}^{m+n} \prod_{h=1}^{r_k} 
(\alpha(g_{h,1}^{(k)}) \dots \alpha(g_{h,k}^{(k)})) ,
\label{action alpha gamma}
\end{equation}
and the stabilizer condition is solved by
\begin{equation}
\alpha (g_{h,k}^{(k)}) = g_{\sigma(h), \tau_h (k)}^{(k)} \,, \qquad
\sigma \in S_{r_k} \,, \quad \tau_h \in \bb{Z}_k \,, \quad
(h=1,2,\dots, r_k).
\label{action alpha gamma2}
\end{equation}
Thus, for each $\gamma$, $\alpha$ should belong to the direct product of the wreath product groups,
\begin{equation}
\alpha \in \prod_{k=1}^{m+n} S_{r_k} [ \bb{Z}_k ] \equiv {\rm Stab} (\gamma), \qquad
\Bigl| {\rm Stab} (\gamma) \Bigr| = \prod_{k=1}^{m+n} k^{r_k} \, r_k! \,.
\label{def:stab gamma}
\end{equation}
The symbol $|G|$ means the order of the group $G$.

The number of permutations in $S_m$ with the cycle structure $p \vdash m$ is given by the orbit-stabilizer theorem, 
\begin{equation}
|T_p| = \frac{|S_m|}{\prod_{k=1}^m |S_{p_k} [ \bb{Z}_k]|} = \frac{m!}{\prod_k k^{p_k} \, p_k!} \,.
\label{orbit stabilizer}
\end{equation}
We can rewrite $N^{\rm MT}_{m,n}$ in \eqref{def:NSU2 mn} as
\begin{equation}
\begin{aligned}
N^{\rm MT}_{m,n} = \frac{1}{m! \, n!} \sum_{\substack{p \vdash m \\ q \vdash n}} 
|T_p| \, |T_q| \, \Bigl| {\rm Stab} (\gamma) _{\gamma \in T_p \times T_q} \Bigr| 
= \sum_{\substack{p \vdash m \\ q \vdash n}} \prod_k \frac{(p_k+q_k)!}{p_k! \, q_k!} \,.
\end{aligned}
\label{NSU2 mn-2}
\end{equation}

Consider the generating function \eqref{def:Z0mt}.
The double sum $\sum_m \sum_{p \vdash m}$ can be transformed to an infinite product $\prod_{k=1}^\infty \sum_{p_k = 0}^{\infty}$\,, and thus
\begin{equation}
Z_0^{\rm MT}(x,y) = \prod_{k=1}^\infty \sum_{p_k = 0}^{\infty} \sum_{q_k = 0}^{\infty} \frac{(p_k+q_k)!}{p_k! q_k!} x^{k p_k} y^{k q_k}
= \prod_{k=1}^\infty \sum_{r_k = 0}^{\infty} (x^k + y^k)^{r_k}
= \prod_{k=1}^\infty \frac{1}{1- x^k - y^k} \,.
\label{Z0mt-2}
\end{equation}
The first few terms read
\begin{multline}
Z_0^{\rm MT}(x,y) = 1+  (x+y)+2  \left(x^2+x y+y^2\right)
+ \left(3 x^3+4 x^2 y+4 x y^2+3 y^3\right)
\\
+ \left(5 x^4+7 x^3 y+10 x^2 y^2+7 x y^3+5 y^4\right)  
+ \dots .
\label{Z0mt-2 expand}
\end{multline}
The series gives the number of multi-trace operators in the $SU(2)$ sector of $\cN=4$ SYM with $U(N_c)$ gauge group. For $SU(N_c)$ theories, we subtract the terms with $p_1 > 0$ or $q_1 > 0$ in \eqref{Z0mt-2}, which gives
\begin{equation}
\tilde Z_0^{\rm MT} (x,y) 
= (1-x)(1-y) \prod_{k=1}^\infty \frac{1}{1- x^k - y^k} \,.
\label{Z0mt-3}
\end{equation}

At finite $N_c$, fewer terms contribute to the generating function \eqref{def:Z0mt}, which modifies the expansion \eqref{Z0mt-2 expand}. The precise expression will be reviewed in Section \ref{sec:finite Nc}. Our formula \eqref{Z0mt-2} is valid up to the order $x^m y^n$ with $m+n \le N_c$\,.

\subsection{Power enumeration theorem}\label{sec:PET}

We review another derivation of the tree-level generating function based on P\'olya Enumeration Theorem \cite{Sundborg:1999ue}.

Consider a single-trace operator with length $p$. We define the domain $D = \{1,2,\dots, p\}$ and the range $R = \{Z, W\}$. A single-trace operator is graphically equivalent to a necklace, that is the map $D \to R$ modulo the action of the cyclic group $\bb{Z}_p$ acting on $D$,
\begin{equation}
\text{Single-trace operator} \ \ \leftrightarrow \ \ 
\text{Necklace} = {\rm Map} \( \bb{Z}_p \backslash D \to R \) .
\label{ST to graph}
\end{equation}
We associate the weights in $R$ by $c(x,y) = x + y$, where $x^m y^n$ corresponds to the operator $W^m Z^n$. Then, P\'olya Enumeration Theorem says that the generating function of the number of graphs \eqref{ST to graph} is given by
\begin{equation}
Z_0^{\rm ST} (x, y) =
\sum_{p} Z_{\bb{Z}_p} \Big( c(x,y), c(x^2,y^2), \dots , c(x^p,y^p) \Big),
\label{Z0st Polya}
\end{equation}
Here $Z_{\bb{Z}_p} (s_1, s_2, \dots , s_p)$ is the cycle index of the cyclic group,
\begin{equation}
Z_{\bb{Z}_p} (s_1, s_2, \dots, s_p) = \frac{1}{p} \sum_{h|p} {\rm Tot} (h) \, s_h^{p/h} \,,
\label{def:cycle index}
\end{equation}
where we take a sum over $h$ such that $p/h$ is a positive integer, and ${\rm Tot} (h)$ is Euler's totient function defined by
\begin{equation}
{\rm Tot} (h) = \sum_{d=1}^h \delta ({\rm gcd}(d,h), 1).
\label{def:Totient}
\end{equation}
By combining \eqref{Z0st Polya} with \eqref{def:cycle index} and writing $p=hs$, we get
\begin{equation}
Z_0^{\rm ST} (x, y) =
\sum_{h} \sum_{s} {\rm Tot} (h) \, \frac{( x^h + y^h )^{s}}{hs}
= - \sum_{h} \frac{{\rm Tot} (h)}{h} \, \log \(1 -  x^h - y^h \).
\label{Z0st Polya-2}
\end{equation}
The first few terms read
\begin{multline}
Z_0^{\rm ST} (x, y) = (x+y)
+ \left(x^2+x y+y^2\right)+ \left(x^3+x^2 y+x y^2+y^3\right)
\\[1mm]
+ \left(x^4+x^3 y+2 x^2 y^2+x y^3+y^4\right)
+ \left(x^5+x^4 y+2 x^3 y^2+2 x^2 y^3+x y^4+y^5\right)
+ \dots .
\end{multline}

The generating function of multi-trace operator is given by the plethystic exponential of the single-trace generating function,
\begin{equation}
Z_0^{\rm MT} (x, y) = \exp \Bigg( \sum_{m=1} \frac{Z_0^{\rm ST} (x^m, y^m)}{m} \Bigg)
= \prod_{d=1} \frac{1}{1 -  x^d - y^d} \,,
\end{equation}
where we used $\sum_{j|d} {\rm Tot} (j) = d$. This result agrees with \eqref{Z0mt-2}.
For $SU(N_c)$ theories, we subtract the $p=1$ term in \eqref{Z0st Polya}, 
\begin{equation}
\tilde Z_0^{\rm MT} (x,y) 
= \exp\( - \sum_{n=1}^\infty \frac{x^n + y^n}{n} \) \prod_{k=1}^\infty \frac{1}{1- x^k - y^k} \,.
\label{Z0mt Polya-2}
\end{equation}
in agreement with \eqref{Z0mt-3}.

\subsection{Counting single-traces}\label{sec:count ST}

For later purposes, we rederive the generating function of the number of single-trace operators by counting the solutions \eqref{def:NSU2 mn} under the constraint $\alpha \in \bb{Z}_L$\,, with $L=m+n$.

We will obtain
\begin{equation}
N_{m,n}^{\rm ST} = \begin{cases}
\ds \sum_{\substack{d=1 \\ d | m, \ d | n}}^L \frac{(L/d)!}{(m/d)! (n/d)!} \, \frac{{\rm Tot} (d)}{L} &\qquad (m \neq 0, L)
\\[3mm]
1 &\qquad (m=0, L),
\end{cases}
\label{number STmn}
\end{equation}
which is derived as follows.
Suppose $m$ and $n$ are divisible by a positive integer $d$,
\begin{equation}
(m,n, L) = (d m', d n', d \ell), \qquad
m' + n' = \ell, \qquad
(d=1,2, \dots, L).
\label{def:dmn}
\end{equation}
The upper bound of $d$ is $L$ if $mn=0$, and ${\rm Min} (m,n)$ otherwise.
Choose $\tilde \mu \in \bb{Z}_d^{\ell} = \bb{Z}_d^{m'} \times \bb{Z}_d^{n'}$ from $S_m \times S_n$ and write $\alpha \in \bb{Z}_L$ as
\begin{equation}
\begin{gathered}
\alpha = \Bigl( a_1 \dots a_\ell \ \tilde \mu^\kappa (a_1) \dots \tilde \mu^\kappa (a_\ell) \, \tilde \mu^{2\kappa} (a_1) \dots \tilde \mu^{2\kappa} (a_\ell) \dots
\tilde \mu^{(d-1)\kappa}(a_1) \dots \tilde \mu^{(d-1)\kappa} (a_\ell) \Bigr) ,
\\
1 \le \kappa < d, \qquad {\rm gcd} (\kappa, d) =1.
\end{gathered}
\label{sol:stabilizer st}
\end{equation}
This set of $(\tilde \mu, \kappa, \alpha)$ is the general solution to the conditions
\begin{equation}
\alpha = \tilde \mu \alpha \tilde \mu^{-1} \qquad {\rm and} \qquad
\alpha \in \bb{Z}_L \,.
\end{equation}

Let us parametrize $\mu = \tilde \mu$ as
\begin{equation}
\tilde \mu = \begin{bmatrix}
(\tilde m_{11} \, \tilde m_{12} \dots \tilde m_{1d}) \\
\vdots \\
(\tilde m_{m' 1} \, \tilde m_{m' 2} \dots \tilde m_{m' d}) \\[3mm]
(\tilde m_{m'+1,1} \, \tilde m_{m'+1,2} \dots \tilde m_{m'+1,d}) \\
\vdots \\
(\tilde m_{\ell 1} \, \tilde m_{\ell 2} \dots \tilde m_{\ell d}) \\
\end{bmatrix} \quad \in \quad  \bb{Z}_d^{m'} \times \bb{Z}_d^{n'} \,.
\label{parameter tilde mu}
\end{equation}
The number of possible $\tilde \mu$ chosen from $S_{dm'} \times S_{dn'}$ is\footnote{The condition $\tilde \mu \in \bb{Z}_d^{m'} \times \bb{Z}_d^{n'}$ is equivalent to $\tilde \mu \in T_{[d^{m'}]} \times T_{[d^{n'}]}$, and the order of the latter group is given by the orbit-stabilizer theorem \eqref{orbit stabilizer}.}
\begin{equation}
\begin{cases}
\ds \frac{(d m')!}{d^{m'} \, m'!} \, \frac{(d n')!}{d^{n'} \, n'!} &\qquad (m', n'>0) 
\\[3mm]
(\ell-1)! &\qquad (m'n'=0).
\end{cases}
\label{Number tilde mu}
\end{equation}

For each $\tilde \mu$\,, we sum over $\alpha$ as parametrized in \eqref{sol:stabilizer st}. For this purpose we identify $\{ a_1 \,, \dots \,, a_\ell \}$ with some of $\{ \tilde m_{hk} \}$ in \eqref{parameter tilde mu}.
In order to avoid double counting, we fix $a_1=\tilde m_{11}$ and choose 
\begin{equation}
a_h = \tilde m_{\sigma (h) k_h} \qquad 
\( \sigma \in S_{\ell-1} \,, \ 1 \le k_h \le d \) \qquad
\text{for each} \ \ 2 \le h \le d ,
\end{equation}
The number of choices of $a_2 \dots a_\ell$ is\footnote{In other words, we remove the redundancy coming from the overall translation of $\alpha \in \bb{Z}_L$\,.}
\begin{equation}
d^{\ell-1} \, (\ell-1)! \,.
\end{equation}
The number of possible $\kappa$ is ${\rm Tot} (d)$.
Thus, the number of possible $\alpha$, divided by $|S_m \times S_n|$ is
\begin{equation}
\frac{1}{|S_m \times S_n|} \sum_{\mu \in S_m \times S_n} \sum_{\alpha \in T_{[m+n]}} \delta_{m+n} \( \alpha^{-1} \mu \alpha \mu^{-1} \)
= \sum_{\substack{d=1 \\ d|m, \ d|L}}^L \frac{(m'+n')!}{m'! \, n'!} \, \frac{{\rm Tot} (d)}{L} \,,
\label{num stabilizer st}
\end{equation}
which is \eqref{number STmn}. 
This result is formally correct when $m'n'=0$ thanks to $\sum_{d|L} {\rm Tot}(d) = L$.

Therefore, the tree-level generating function is given by
\begin{equation}
Z_0^{\rm ST} (x,y) 
= \sum_{L} \sum_{m=0}^{L}
\sum_{\substack{d=1 \\ d | m, \ d | L}}^L x^m y^n \, \frac{(L/d)!}{(m/d)! ((L-m)/d)!} \, \frac{{\rm Tot} (d)}{L} .
\end{equation}
To simplify it, we apply the formulae
\begin{equation}
\begin{aligned}
\sum_{L}^\infty \sum_{m=0}^L \sum_{\substack{d=1 \\ d|m, \ d|L}}^{L-1} f_d (m,L-m)
&=
\sum_{d=1}^\infty \sum_{L}^\infty \sum_{m=0}^L
f_d (dm, d(L-m) ) ,
\\
\sum_{m=0}^{L} x^{dm} y^{d(L-m)} \, \frac{L!}{m! (L-m)!} 
&= (x^d + y^d)^L ,
\end{aligned}
\label{double sum formula}
\end{equation}
to obtain
\begin{equation}
Z_0^{\rm ST} (x,y) = \sum_{d=1}^\infty \sum_{L}  \frac{{\rm Tot} (d)}{d} \frac{(x^d + y^d)^L}{L} \,,
\end{equation}
which is \eqref{Z0st Polya-2}.

\bigskip
Consider an example. If $(m,n)=(3,3)$ and $d=3$, we find
\begin{equation}
S_3 \times S_3 \supset \bb{Z}_3^1 \times \bb{Z}_3^1 \ni \tilde \mu = \pare{
\begin{bmatrix}
(123) \\ (456) 
\end{bmatrix}, \ 
\begin{bmatrix}
(132) \\ (456) 
\end{bmatrix}, \ 
\begin{bmatrix}
(123) \\ (465) 
\end{bmatrix}, \ 
\begin{bmatrix}
(132) \\ (465) 
\end{bmatrix}
},
\end{equation}
which is consistent with $\{ 3!/(3^1 \, 1!) \}^2 = 4$ in \eqref{Number tilde mu}. For each $\tilde \mu$ we generate 
\begin{equation}
\alpha = (a_1 \, a_2 \, \tilde \mu^\kappa (a_1) \, \tilde \mu^\kappa (a_2) \, \tilde \mu^{2\kappa} (a_1) \, \tilde \mu^{2\kappa} (a_2) ), \qquad (\kappa=1,2).
\end{equation}
The possible choices of $(a_1, a_2)$ are $(1,4), (1,5), (1,6)$. Thus
\begin{equation}
\frac{1}{|S_3 \times S_3|} \sum_{\mu \in \bb{Z}_3 \times \bb{Z}_3} \sum_{\alpha \in T_{[6]}} \delta_6 \( \alpha^{-1} \mu \alpha \mu^{-1} \)
= \frac{4 \times 2 \times 3}{3!^2}
= \frac{2}{3} \,,
\end{equation}
which agrees with \eqref{num stabilizer st}.

\subsection{Partition function at finite $N_c$}\label{sec:finite Nc}

The exact tree-level partition function of $\cN=4$ SYM at finite $N_c$ can be computed precisely in various methods. We briefly review these arguments, to understand the finite $N_c$ corrections in \eqref{cZ finite Nc}.

A straightforward method is to compute the grand partition function is to evaluate the path integral of $\cN=4$ SYM action \cite{Yamada:2006rx}. Let $a_0$ be the zero-mode of the gauge field $A_0$ and $U = \exp(i \beta a_0)$. The grand partition function in the complete $PSU(2,2|4)$ sector is
\begin{equation}
\cZ^{\rm complete}_{N_c} (w) = \int dU_{SU(N_c)} \, \exp \( \frac{1}{n} \sum_{n=1}^\infty \pare{
\zeta_B (w^n) + (-1)^{n+1} \zeta_F (w^n) }
\tr_{\! {\rm adj}} \( U^n \) \) 
\label{complete Z}
\end{equation}
where $\zeta_B (w), \zeta_F (w)$ are functions of chemical potentials, and $dU_{SU(N_c)}$ is the $SU(N_c)$ Haar measure.\footnote{This is the result for $SU(N_c)$ gauge group. For $U(N_c)$, we replace $\tr_{\! {\rm adj}} \( U^n \)$ by $\tr_{\! {\rm adj}} \( U^n \) + 1$. Recall that the $U(1)$ part of $\cN=4$ SYM is free since all interactions are of commutator type.}

The complete partition function \eqref{complete Z} can be reduced to the one in the $SU(2)$ sector by setting
\begin{equation}
\zeta_B (w^n) = x^n + y^n, \qquad
\zeta_F (w^n) = 0,
\end{equation}
which gives $\cZ^{\rm exact}_{N_c} (\beta, x,y)$ in \eqref{cZ finite Nc}.
It turns out that the resulting expression is identical to the Molien-Weyl formula which gives the Hilbert-Poincar\'e series of $GL(N_c)$ invariants \cite{Gray:2008yu}.
For example, the Molien-Weyl formula for the gauge group $U(N_c)$ with $q$ variables, corresponding to the $SU(q)$ sector, can be written as\footnote{The explicit form of the Molien-Weyl formula depends on the choice of basis of the (adjoint) representation of $U(N_c)$. The convention of \cite{Djokovic} is used here for efficient evaluation.}
\begin{gather}
\cZ^{SU(q)}_{N_c} (x_1, \dots, x_q) = \frac{1}{(2 \pi i)^{N_c-1}} \, \frac{1}{\prod_{i=1}^q (1-x_i)^n} \,
\oint_U \frac{dt_1}{t_1 } \dots \oint_U \frac{dt_{N_c-1}}{t_{N_c-1} } \, 
\prod_{r=1}^{n-1} \prod_{k=1}^r \frac{\chi_{k,r}^+ (1,t)}{\phi_{k,r} (x,t)} \,,
\notag \\
\chi_{k,r}^\pm (\alpha, t) = 1 - \alpha \, \prod_{j=k}^r t_j^{\pm 1} \,,\qquad
\phi_{k,r} (x,t) = \prod_{\ell=1}^q \chi_{k,r}^+ (x_\ell, t) \, \chi_{k,r}^- (x_\ell , t) ,
\label{Molien-Weyl SUq}
\end{gather}
where $U$ is the counterclockwise contour of unit radius.\footnote{This formula is elaborated further as the highest weight generating function \cite{Hanany:2014dia}.}
It can also be written as \cite{Dolan:2007rq,Brown:2007xh}
\begin{equation}
\cZ^{SU(q)}_{N_c} (x_1, \dots, x_q) = \sum_{L=0} \sum_{R \vdash L} 
\sum_{\substack{\Lambda \vdash L \\[1mm] {\rm Row}(\Lambda) \le q}} C(R,R,\Lambda) s_\Lambda (x_1, \dots x_q),
\label{Z0mt-exact}
\end{equation}
where $C(R,R,\Lambda)$ is the Clebsch-Gordan multiplicity defined by $R \otimes R = \oplus \Lambda^{C(R,R,\Lambda)}$ as $S_L$-modules, and $s_\Lambda (x_1, \dots x_q)$ is the Schur polynomial. We sum over the partitions $\Lambda$ having at most $q$ rows, since $\Lambda$ is related to the $SU(q)$ global symmetry. 
At finite $N_c$ we should sum over the partitions $R$ having at most $N_c$ rows in \eqref{Z0mt-exact}.

One can evaluate the formula \eqref{Molien-Weyl SUq} or \eqref{Z0mt-exact} explicitly when $N_c$ and $q$ are small. At $(N_c,q)=(2,2)$ we obtain
\begin{equation}
\cZ^{SU(2)}_{N_c} (x,y) = (Z_0^{\rm MT})_{N_c=2}^{\rm exact} (x,y) = \frac{1}{(1-x y)} \, \prod_{k=1}^2 \frac{1}{(1-x^k)(1-y^k)} \,,
\label{Nc=2 SU(2) exact}
\end{equation}
in agreement with \cite{Kimura:2009ur} for $q=2$. The formula also reproduces the $q>2$ cases in \cite{Harmark:2014mpa}.

\section{One-loop counting}\label{sec:one-loop count}

We compute the sum of anomalous dimensions at one-loop in the $SU(2)$ sector in two ways, which we call Partition form and Totient form. The corresponding generating function gives the partition function at one-loop.

\subsection{Mixing matrix}

The dilatation operator of $\cN=4$ SYM is given by \cite{Beisert:2003tq,Beisert:2003jj},
\begin{equation}
\fD = \sum_{n=0} \lambda^n \, \fD_{2n} 
= \tr (W \check{W} + Z \check{Z} ) - \frac{2 \lambda}{N_c} : \! \tr [W,Z] [\check{W}, \check{Z} ] \! :
+ O(\lambda^2).
\end{equation}
Let $\cH_{m,n}$ be the Hilbert space of all gauge-invariant operators in the $SU(2)$ sector with the R-charges $(m,n)$. We define the mixing matrix as
\begin{equation}
\fD_2 \, \cO_\alpha \equiv \frac{2}{N_c} \, (M_2)_\alpha {}^\beta \, \cO_\beta \,.
\label{def:mixing matrix}
\end{equation}
On the permutation basis introduced in Section \ref{sec:pm basis}, the mixing matrix inside $\cH_{m,n}$ takes the form \cite{Bellucci:2004ru}
\begin{equation}
\begin{aligned}
(M_2)_{\alpha}{}^{\beta} &= \sum_{i \neq j}^L \Big[
\delta_L ( [\beta^{-1}] [\alpha] \spar{i \alpha(j)} ) - \delta_L ( [\beta^{-1}] (ij) [\alpha] (ij) \spar{i \alpha(j)} )
\Big] ,
\\
&= \frac{1}{m! n!} \sum_{i \neq j}^L \sum_{\mu \in S_m \times S_n}
\delta_L \Bigl( \mu \beta^{-1} \mu^{-1} \Bigl\{ \alpha - (ij) \alpha (ij) \Bigr\} \spar{i \alpha(j)} \Bigr).
\end{aligned}
\label{M2 ab}
\end{equation}
where we introduced the notation $L=m+n$,
\begin{equation}
\spar{ij} = \begin{cases}
(ij) &\qquad (i \neq j) \\
N_c &\qquad (i = j),
\end{cases}
\label{def:double bracket}
\end{equation}
and denoted the equivalence class by
\begin{equation}
[\alpha] = \frac{1}{|S_m \times S_n|} \, \sum_{\gamma \in S_m \times S_n} \gamma \alpha \gamma^{-1} .
\label{def:equiv class}
\end{equation}

We will evaluate the sum of one-loop dimensions at a fixed $(m,n)$,
\begin{equation}
\vev{ M_2 }_{m,n} = \sum_{\alpha, \beta \in \cH_{m,n}} (M_2)_{\alpha}{}^{\beta} \, \delta_\beta{}^{\alpha} \,.
\label{def:average M2mn}
\end{equation}
Since the gauge-invariant operator is in one-to-one correspondence with the equivalence class \eqref{def:equiv class}, we can rewrite the sum
\eqref{def:average M2mn} as
\begin{equation}
\vev{ M_2 }_{m,n} = \frac{1}{(m! \, n!)^2} \sum_{\alpha, \beta \in S_{L}} \sum_{\gamma_1, \gamma_2 \in S_m \times S_n}
(M_2)_{\gamma_1 \alpha \gamma_1^{-1}}{}^{\gamma_2 \beta \gamma_2^{-1}} \, 
\delta_{L} \( \gamma_2 \beta^{-1} \gamma_2^{-1} \gamma_1 \alpha \gamma_1^{-1} \) .
\end{equation}
According to \eqref{M2 ab}, the mixing matrix is invariant inside the same conjugacy class. By writing $\gamma_2^{-1} \gamma_1 \equiv \gamma$, we find 
\begin{equation}
\begin{aligned}
\vev{ M_2 }_{m,n} &= \frac{1}{m! \, n!} \sum_{\alpha, \beta \in S_{L}} \sum_{\gamma \in S_m \times S_n}
(M_2)_{\alpha}{}^{\beta} \, 
\delta_{L} \( \beta^{-1} \gamma \alpha \gamma^{-1} \) ,
\\ 
&= \frac{1}{m! \, n!} \sum_{\alpha \in S_{L}} \sum_{\gamma \in S_m \times S_n} (M_2)_{\alpha}{}^{\gamma \alpha \gamma^{-1}} \,,
\\
&= \sum_{\alpha \in S_{L}} (M_2)_{\alpha}{}^{\alpha} \,,
\\
&= \frac{1}{m! \, n!} \sum_{i \neq j}^{L} 
\sum_{\alpha \in S_{L}} \sum_{\mu \in S_m \times S_n} 
\delta_{L} \Big( \mu \alpha^{-1} \mu^{-1} \pare{ \alpha - (ij) \alpha (ij) } \spar{ i \alpha(j) } \Big).
\end{aligned}
\label{average M2mn}
\end{equation}

Let us inspect the argument of the $\delta$-function.
Recall that any permutation can be decomposed into the product of transpositions, like $(1234)=(34)(23)(12)$. The number of transpositions defines the parity of a permutation, which is conserved at any orders of perturbative $1/N_c$ expansion.\footnote{Conversely said, finite $N_c$ constraints mix permutations with different parity.} In particular, odd powers of transpositions cannot become the identity, and only the planar term $i=\alpha(j)$ contributes in \eqref{average M2mn}. Thus,
\begin{multline}
\vev{M_2}_{m,n} = \frac{N_c}{m! \, n!} \sum_{i \neq j}^L  
\sum_{\alpha \in S_L} \ \sum_{\mu \in S_m \times S_n} 
\delta_L (i \alpha(j)) \ \times
\\
\pare{
\delta_L ( \mu \alpha^{-1} \mu^{-1} \alpha ) - 
\delta_L \Big( \mu \alpha^{-1} \mu^{-1} (ij) \alpha (ij) \Big) 
},
\label{su2 1loop av mn}
\end{multline}
where $L = m+n$.
The generating function of the sum of one-loop dimensions is defined by
\begin{equation}
Z_2^{\rm MT} (x, y) \equiv \sum_{m,n=0}^\infty \vev{M_2}_{m,n} \, x^m y^n .
\end{equation}

\subsection{Partition form}\label{sec:Partition form}

We evaluate the sum of dimensions \eqref{su2 1loop av mn} by generalizing the methods used in Section \ref{sec:tree partition}.

\subsubsection{First term}\label{sec:part 1st}

Consider the first term of \eqref{su2 1loop av mn},
\begin{equation}
\vev{M_2}_{m,n}^{\rm (1st)} 
= \frac{N_c}{m! \, n!} \sum_{i \neq j}^L 
\sum_{\alpha \in S_L} \sum_{\mu \in S_m \times S_n} 
\delta_L (i \alpha(j)) \ 
\delta_L ( \mu \alpha^{-1} \mu^{-1} \alpha ) .
\label{def:M2SU2 1st}
\end{equation}
We denote the cycle type of $\mu$ by $p \vdash m$, $q \vdash n$ and define $r_k = p_k + q_k$\,. We parametrize $\mu$ by $\mu = \prod_{k=1}^{m+n} \prod_{h=1}^{r_k} \( m^{(k)}_{h,1} m^{(k)}_{h,2} \dots m^{(k)}_{h,k} \)$ as in \eqref{parametrize gamma}.
The condition $\mu \alpha^{-1} \mu^{-1} \alpha = 1$ imposes that $\alpha$ should belong to the stabilizer of $\mu$.

Suppose that $i$ and $j$ are part of the cycle of $\mu$ of length-$a$ and length-$b$, respectively. There are $L(L-1)$ choices of $\{ i,j \}$, which can be written as
\begin{equation}
L(L-1) = \sum_{a,b=1}^{L} a b \, r_a r_b - \sum_{a=1}^{L} a r_a
= \sum_{a \neq b} a b \, r_a r_b 
+ \sum_a a r_a (a r_a -1) .
\label{decompose ij}
\end{equation}
From \eqref{action alpha gamma2} we see that $\alpha \in {\rm Stab} (\mu)$ permutes $\pare{1,2,\dots, L}$ only among those having the same cycle length in $\mu$. Thus, the condition $i=\alpha(j)$ results in $a=b$, so we neglect the terms $a \neq b$ in \eqref{decompose ij}.

Define the number of solutions of the two $\delta$-functions in \eqref{def:M2SU2 1st} for a given $(i,j,\mu)$ by
\begin{equation}
N_{sol} (a,\mu) = \sum_{\alpha \in S_{L}}
\delta(i \alpha(j)) \, \delta_{L} ( \mu^{-1} \alpha^{-1} \mu \alpha ) \Big|_{i,j \ \in\  \text{length-$a$ cycle}} \,.
\label{def:Nsol1 am}
\end{equation}
This can be rewritten as
\begin{equation}
N_{sol} (a,\mu) = \sum_{\alpha \in {\rm Stab} (\mu)}
\delta(i \alpha(j)) \, \Big|_{i,j \, \in\,  \text{length-$a$ cycle}} \,,\qquad
{\rm Stab} (\mu) = \prod_{k=1}^{L} S_{r_k} [\bb{Z}_k]\,.
\end{equation}
If we introduce $\alpha = \alpha_0 \, (ij)$, then
\begin{equation}
N_{sol} (a,\mu) = \sum_{\alpha_0 \in {\rm Stab} (\mu)}
\delta(i \alpha_0(i)) \, \Big|_{i \, \in\,  \text{length-$a$ cycle}} \,.
\end{equation} 
Since the group ${\rm Stab} (\mu)$ acts transitively on $S_{r_a} [\bb{Z}_a] \subset S_{a r_a}$\,, the isotropy group satisfies the property
\begin{equation}
(S_{r_a} [\bb{Z}_a] )^{(i)} \equiv \Bigl\{ g(i) = i \, | \, g \in S_{r_a} [\bb{Z}_a] \Bigr\}, \qquad
\Big| (S_{r_a} [\bb{Z}_a] )^{(i)} \Big| = \frac{1}{a r_a} \, \Big| S_{r_a} [\bb{Z}_a] \Big| .
\label{prop isotropy}
\end{equation}
Thus, the number of solutions in \eqref{def:M2SU2 1st} for a given $\mu \in T_p \times T_q$ is
\begin{equation}
\begin{aligned}
N_{sol} (p,q) &\equiv \sum_{i \neq j}^{L} \sum_{\alpha \in S_{L}}
\delta(i \alpha(j)) \, \delta_{L} ( \mu^{-1} \alpha^{-1} \mu \alpha ) 
\\
&= \sum_{a=1}^{L} a r_a \(a r_a - 1 \) N_{sol} (a,\mu) 
\\
&= \sum_{a=1}^{L} \theta_> (r_a) \, \(a r_a - 1 \) \, \Big| {\rm Stab} (\mu) \Big| _{\mu \in T_p \times T_q} \,,
\end{aligned}
\label{def:Nsol1}
\end{equation}
where 
\begin{equation}
\theta_> (x) = \begin{cases}
1 &\quad (x > 0) \\
0 &\quad (x \le 0).
\end{cases}
\end{equation}

Proceeding as in \eqref{Z0mt-2}, we obtain the generating function for the first term.
\begin{align}
Z_2^{\rm (1st)} (x, y) &= N_c \, \sum_{m,n} \sum_{p \vdash m, q \vdash n} |T_p| |T_q| \, \frac{x^m y^n}{m! \, n!} \, N_{sol} (p,q)
\notag \\
&= N_c \, \sum_{m,n} \sum_{p \vdash m, q \vdash n} \,
\prod_{k=1} (x^k)^{p_k} (y^k)^{q_k} \,\frac{ (p_k + q_k) ! }{p_k! \, q_k! }  \sum_a \theta_> (r_a) \(a r_a - 1\)
\notag \\
&= N_c \, \sum_{L=0}^\infty \sum_{r \vdash L} \prod_{k=1} (x^k + y^k)^{r_k} 
\( L - \sum_{a=1}^L \theta_> (r_a) \) .
\label{gen av 1st}
\end{align}

\bigskip
Consider an example. Let $\mu$ be $(1)(2)(3,4)(5,6)(7,8)(9,10,11)$, having $r=[1^2,2^3,3^1] \vdash 11$. Then ${\rm Stab} (\mu) = S_2[\bb{Z}_1] \cdot S_3[\bb{Z}_2] \cdot S_1 [\bb{Z}_3]$, which has the order $2 \times 48 \times 3 = 244$. Choose $i \neq j$ from $\{ 1,2,\dots, 11 \}$. The list $\{ \alpha(j) \, | \, \alpha \in {\rm Stab} (\mu) \}$ for all $j$ is summarized in Table \ref{tab:alpha j}. The list shows that the number of solutions to $i=\alpha(j)$ is precisely given by the formula \eqref{prop isotropy}, like
\begin{alignat}{9}
(i,j) &= (1,2), &\qquad &\#_{sol} = 144 & &= \Big| 1 \cdot S_3[\bb{Z}_2] \cdot S_1 [\bb{Z}_3] \Big| 
\\[1mm]
(i,j) &= (3,4),(3,5),\dots,(7,8), &\qquad &\#_{sol} = 48 & &= \Big| S_2[\bb{Z}_1] \cdot S_2[\bb{Z}_2] \cdot S_1 [\bb{Z}_3] \Big| 
\\[1mm]
(i,j) &= (9,10),(10,11),(9,11), &\qquad &\#_{sol} = 96 & &= \Big| S_2[\bb{Z}_1] \cdot S_3[\bb{Z}_2] \cdot 1 \Big|.
\end{alignat}

\begin{table}[t]
\begin{center}
\begin{tabular}{c|cc|cccccc|ccc}
$j$ & 1 & 2 & 3 & 4 & 5 & 6 & 7 & 8 & 9 & 10 & 11 \\\hline
\multirow{6}{*}{$\alpha(j)$} & \multicolumn{2}{c|}{$1^{144}$} & \multicolumn{6}{c|}{$3^{48}$} & \multicolumn{3}{c}{$9^{96}$} \\
& \multicolumn{2}{c|}{$2^{144}$} & \multicolumn{6}{c|}{$4^{48}$} & \multicolumn{3}{c}{$10^{96}$} \\
& & & \multicolumn{6}{c|}{$5^{48}$} & \multicolumn{3}{c}{$11^{96}$} \\
& & & \multicolumn{6}{c|}{$6^{48}$} & & & \\
& & & \multicolumn{6}{c|}{$7^{48}$} & & & \\
& & & \multicolumn{6}{c|}{$8^{48}$} & & & \\
\end{tabular}
\caption{Table of $\alpha(j)$ for $1 \le j \le 11$ and $\alpha \in S_2[\bb{Z}_1] \cdot S_3[\bb{Z}_2] \cdot S_1 [\bb{Z}_3]$. The symbol $a^b$ means that the number $a$ appears $b$ times.}
\label{tab:alpha j}
\end{center}
\end{table}

\subsubsection{One-loop generating function}\label{sec:compute gn mt}

We will analyze the second term of the mixing matrix in Appendix \ref{app:part 2nd}. Here we summarize the results by combining \eqref{gen av 1st} and \eqref{gen av 2nd}.

The generating function of the sum of one-loop dimensions over all multi-trace operators in the $SU(2)$ sector is given by
\begin{equation}
\begin{aligned}
Z_2^{\rm MT} (x, y) &= N_c \, \sum_{L=0}^\infty \sum_{r \vdash L} \,
\prod_{k=1}^\infty (x^k + y^k)^{r_k} \Big\{ 
L - \sum_{a=1}^L \theta_> (r_a) - \Theta (r) \Big\} ,
\\
&= N_c \, \sum_{L=0}^\infty \sum_{r \vdash L} \,
\prod_{k=1}^\infty (x^k + y^k)^{r_k} \ \Big\{ L - \sum_{a=1}^L \theta_> (r_a)
- \sum_{a=1}^{L/2} a \, (r_a+1) \theta_> (r_{2a}) 
\\
&\qquad
- 2 \sum_{a< b}^L \theta_> (L+1-a-b) \theta_> (r_a) \theta_> (r_b)  
- \sum_{a=1}^{L/2} \theta_> (r_a-1) 
\Big\}.
\end{aligned}
\label{gen av 1-loop SU2}
\end{equation}
The summand can be negative for some $r \vdash L$, though the sum becomes non-negative if we sum over all $r \vdash L$.\footnote{Negative terms are needed to kill the coefficients of BPS terms.} The first few terms read
\begin{multline}
\frac{Z_2^{\rm MT} (x, y)}{N_c} = 
6 x^2 y^2 + \left(10 x^3 y^2+10 x^2 y^3\right)
+ \left(26 x^4 y^2+36 x^3 y^3+26 x^2 y^4\right)
\\
+ \left(44 x^5 y^2+84 x^4 y^3+84 x^3 y^4+44 x^2 y^5\right)
+ \left(84 x^6 y^2+176 x^5 y^3+254 x^4 y^4+176 x^3 y^5+84 x^2 y^6\right)
\\
+ \left(134 x^7 y^2+348 x^6 y^3+548 x^5 y^4+548 x^4 y^5+348 x^3 y^6+134 x^2 y^7\right)
+ \dots   
\label{gen av 1-loop SU2 exp}
\end{multline}
The term $6 x^2 y^2$ is responsible for the one-loop dimensions of the $SU(2)$ Konishi descendant, which is $3 N_c \, g_{\rm YM}^2/(4 \pi^2)$.

\subsection{Totient form}\label{sec:compute gn st}

We compute the generating function of the sum of one-loop dimensions in another way. First, we compute the one-loop generating function for single-trace operators by imposing $\alpha \in \bb{Z}_L$\,, as done in Section \ref{sec:count ST}. 
Then, we conjecture the generating function for multi-traces, by writing the plethystic exponential of the single-trace results.

\subsubsection{First term}

Let $d \ge 1$ be a divisor of $m$ and $n$ as in \eqref{def:dmn}.
Specify the cycle type of $\mu$ to $p= [d^{m'}]$ and $q = [d^{n'}]$ and $\alpha$ to $[L]$ simultaneously. Consider the first term of the one-loop mixing matrix:
\begin{equation}
\vev{M_2}_{m,n}^{\rm (1st)} 
= \frac{N_c}{m! \, n!} \sum_{i \neq j}^{L} 
\sum_{\alpha \in T_{[L]}} \sum_{\substack{d=1 \\ d|m, \ d|L}}^L
\sum_{\mu \in T_{[d^{m'}]} \times T_{[d^{n'}]}} 
\delta(i \alpha(j)) \ 
\delta_{L} ( \alpha^{-1} \mu \alpha \mu^{-1} ) .
\end{equation}
The number of solutions to the stabilizer condition $\alpha =  \mu \alpha \mu^{-1}$ is given by \eqref{num stabilizer st}. 
For each $(j, \mu = \tilde \mu)$ and $\alpha$ given by \eqref{sol:stabilizer st}, there is only one $i \in \pare{1,2,\dots, L}$ satisfying $i =\alpha (j)$. Hence the sum over $i,j$ gives a factor of $L$, leading to 
\begin{equation}
\vev{M_2}_{m,n}^{\rm (1st)} 
= N_c \sum_{\substack{d=1 \\ d|m, \ d|L}}^L \frac{(m'+n')!}{m'! \, n'!} \, {\rm Tot} (d) .
\label{gen 1L-ST 1st}
\end{equation}
Note that $d=L$ is possible only when $mn=0$.

\subsubsection{One-loop generating function}

The second term of the one-loop mixing will be computed in Appendix \ref{app:totient 2nd}. By combining the results \eqref{gen 1L-ST 1st} and \eqref{gen 1L-ST 2nd}, we obtain the one-loop generating function for single-trace operators as\footnote{The sum over $L$ begins with $L=2$, because $L=1$ are BPS operators.}
\begin{align}
Z_2^{\rm ST} (x,y) &= N_c \sum_{L=2}^\infty \sum_{m=0}^L x^m y^n \, \Biggl(
\sum_{\substack{d=1 \\ d|m, \ d|L}}^L \frac{(L/d)!}{(m/d)! \, (n/d)!} \, {\rm Tot} (d)
\notag \\
&\hspace{5mm} 
- \( 1 - \delta(mn,0) \) 2 \, \delta({\rm gcd}(m,n), 1) 
- \delta(mn,0) \, {\rm Tot} (L)
\notag \\
&\hspace{10mm} 
- \sum_{\substack{d=1 \\ d|m, \ d|L}}^{L-1} {\rm Tot} (d)  \pare{
\frac{(L/d-2)!}{(m/d-2)! (n/d)!} + \frac{(L/d-2)!}{(m/d)! (n/d-2)!} } \Biggr) ,
\end{align}
with $n=L-m$.
On the first line, $d=L$ is possible only if $mn=0$. This contribution is cancelled exactly by the last term on the second line. It follows that
\begin{align}
\frac{Z_2^{\rm ST} (x,y)}{N_c} &= - 2 \sum_{L=2}^\infty \sum_{m=1}^{L-1} x^m y^n \, \delta({\rm gcd}(m,n), 1) 
\\
&\hspace{5mm} 
+ \sum_{L=2}^\infty \sum_{m=0}^L x^m y^n \sum_{\substack{d=1 \\ d|m, \ d|L}}^{L-1} 
\frac{(L/d)!}{(m/d)! \, (n/d)!} \, {\rm Tot} (d) 
\pare{ 1 - \frac{m (m/d-1)}{L (L/d-1)} - \frac{n (n/d-1)}{L (L/d-1)} } .
\notag 
\end{align}
We apply the formula \eqref{double sum formula} and simplify the second line as
\begin{align}
\frac{Z_2^{\rm ST} (x,y)}{N_c} &= - 2 \sum_{L=2}^\infty \sum_{m=1}^{L-1} x^m y^{L-m} \, \delta({\rm gcd}(m,L), 1) 
\\
&\hspace{-10mm} 
+ \sum_{d=1}^\infty {\rm Tot} (d) \sum_{L=2}^\infty \sum_{m=0}^L x^{dm} y^{d(L-m)} \, 
\frac{L!}{m! \, (L-m)!} \, 
\pare{ 1 - \frac{m (m-1)}{L (L-1)} - \frac{(L-m) (L-m-1)}{L (L-1)} } ,
\notag \\[2mm]
&= - 2 \sum_{L=2}^\infty \sum_{m=1}^{L-1} x^m y^{L-m} \, \delta({\rm gcd}(m,L), 1) 
+ 2 \sum_{d=1}^\infty {\rm Tot} (d) \sum_{L=2}^\infty x^d y^d (x^d + y^d)^{L-2} ,
\notag \\[2mm]
&= - 2 \sum_{L=2}^\infty \sum_{m=1}^{L-1} x^m y^{L-m} \, \delta({\rm gcd}(m,L), 1) 
+ 2 \sum_{d=1}^\infty {\rm Tot} (d) \frac{x^d y^d}{1- x^d - y^d} \,.
\label{Z2st formula}
\end{align}
The first few terms are
\begin{align}
\frac{Z_2^{\rm ST} (x,y)}{N_c} &= 6 x^2 y^2 
+4 x^2 y^2 (x+y)
+2 x^2 y^2 \left(5 x^2+8 x y+5 y^2\right)
+2 x^2 y^2 \left(4 x^3+9 x^2 y+9 x y^2+4 y^3\right)
\notag \\
&\quad +2 x^2 y^2 \left(7 x^4+14 x^3 y+24 x^2 y^2+14 x y^3+7 y^4\right)
+ \dots .
\end{align}

In the degeneration limit $x=y=z$, the generation function \eqref{Z2st formula} becomes
\begin{equation}
\begin{aligned}
\frac{Z_2^{\rm ST} (z)}{N_c} &= - 2  \sum_{L=2}^\infty z^L \, {\rm Tot} (L)
+  \sum_{d=1}^\infty {\rm Tot} (d) \frac{2 z^{2d}}{1 - 2 z^d} \,,
\\[1mm]
&= 2  \pare{ z - \sum_{L=1}^\infty {\rm Tot} (L) \, z^L \, \frac{1 - 3 z^L}{1 - 2 z^L} },
\end{aligned}
\label{Z2st formula deg}
\end{equation}
in perfect agreement with \cite{Spradlin:2004pp}.

\bigskip
We conjecture that the one-loop generating function for multi-traces is given by the plethystic exponential of the singe-trace results \eqref{Z2st formula}:
\begin{align}
&Z_2^{\rm MT} (x,y) = Z_0^{\rm MT} (x,y)  \sum_{k=1}^\infty Z_2^{\rm ST} (x^k,y^k) ,
\label{Z2mt formula} \\
& = 2 N_c \prod_{h=1}^\infty \frac{1}{1- x^h - y^h} \sum_{k=1}^\infty 
\( \sum_{d=1}^\infty {\rm Tot} (d) \frac{x^{kd} y^{kd}}{1- x^{kd} - y^{kd}}
- \sum_{L=2}^\infty \sum_{m=1}^{L-1} x^{km} y^{k(L-m)} \, \delta({\rm gcd}(m,L), 1) \).
\notag
\end{align}
One can check that its expansion in small $x, y$ agrees with \eqref{gen av 1-loop SU2 exp}.
The first line is generalization of the single-variable case discussed in \cite{Spradlin:2004pp}.

\subsection{Comparison with Bethe Ansatz}\label{sec:BAE}

We compare our results with the prediction of Bethe Ansatz Equations (BAEs) for XXX${}_\frac12$ spin chain.
The single-trace operators of $\cN=4$ SYM in the $SU(2)$ sector correspond to the level-matched and physical solutions of BAE. 
The Bethe roots of the level-matched solutions satisfy
\begin{equation}
\bb{Q} (i/2) = \bb{Q} (-i/2), \qquad \bb{Q} (v) = \prod_{j} (v-u_j),
\end{equation}
and in the physical solutions the second Q-function $\tilde{\bb{Q}}(v)$ must be a polynomial in $v$, as clarified in \cite{Hao:2013jqa,Marboe:2016yyn}.

A solution of BAEs is called regular if no Bethe roots are located at infinity.
Regular BAE solutions correspond to the $SU(2)$ highest weight states. If we denote a state with $W^m Z^n$ by $\ket{m,n}$, the highest weight states satisfy
\begin{equation}
{\bf S}_- \ket{m,n}_{\rm HWS} = 0, \qquad
{\bf S}_+ \ket{m,n} = \ket{m+1,n-1} ,
\end{equation}
where $\{ {\bf S}_\pm, {\bf S}_3 \}$ are the $SU(2)$ generators.
We also need $m \ge n$ to count the BAE solutions correctly.

We should include exceptional solutions whose energy superficially diverges due to the Bethe roots at $v=\pm i/2$. In such cases, we should regularize the BAEs by introducing twists and by carefully taking the zero-twist limit.
The results are shown in Table \ref{tab:su2 averages}.

\begin{table}[t]
\begin{center}
\begin{tabular}{c|ccccc}
\diagbox{$n$}{$m$} & 1 & 2 & 3 & 4 & 5 \\\hline
1 & 0 & 0 & 0 & 0 & 0\\
2 & & 6 & 10 & 26 & 44\\
3 & & & 36 & 84 & 176
\end{tabular}
\hspace{15mm}
\begin{tabular}{c|ccccc}
\diagbox{$n$}{$m$} & 1 & 2 & 3 & 4 & 5 \\\hline
1 & 0 & 0 & 0 & 0 & 0 \\
2 &  & 6 & 4 & 10 & 8 \\
3 &  &  & 6 & 10 & 14
\end{tabular}
\caption{Average $SU(2)$ one-loop dimensions for $W^m Z^n$ with $m \ge n$, in the unit of $N_c \, g_{\rm YM}^2/(8 \pi^2)$. Left Table shows the sum over all $U(N_c)$ multi-trace operators, and Right Table shows the sum over all single-trace $SU(2)$ highest weight states.}
\label{tab:su2 averages}
\end{center}
\end{table}

A bit of arithmetic is needed to compare the two sets of numbers in Table \ref{tab:su2 averages}.
First, consider the second row. At $(m,n)=(3,2)$ we have $10=4+6$, where 6 comes from the multi-trace $(m,n)=(2,2)+(1,0)$.
At $(m,n)=(4,2)$, we have
\begin{equation}
\begin{aligned}
26 &= 10 + (4 + 6 + 6), \\
\text{Multi-traces} \quad (4,2) &= \pare{ (3,2)+(1,0), (2,2)+(2,0), (2,2)+(1,0)+(1,0) }
\end{aligned}
\end{equation}
Next, consider the third row. At $(m,n)=(3,3)$ we have
\begin{equation}
\begin{aligned}
36 &= 6 + (10) + (2 \times 6+4+4)  \\
{\rm Descendants} \quad (4,2) &\to (3,3) \\
\text{Multi-traces} \quad (3,3) &= \pare{ (2,2) + (1,1) \dots, (3,2)+(0,1), (2,3)+(1,0) } ,
\end{aligned}
\end{equation}
where $\dots$ means other possible partitions, namely $(1,1)$ or $(1,0)+(0,1)$.
At $(m,n)=(4,3)$ we have
\begin{equation}
\begin{aligned}
84 &= 10 + (8+10) + (10 + 6 + 2 \times 4 + 2 \times 4 + 4 \times 6) \\
{\rm Descendants} \quad &\pare{ (5,2) \to (4,3), (4,2) \to (3,3) + (1,0) } \\
\text{Multi-traces} \quad &\bigl\{ (4,2)+(0,1), (3,3)+(1,0), (3,2)+(1,1) \dots, \\
&\hspace{52mm} (2,3) + (1,1) \dots,  (2,2) + (2,1) \dots \bigr\} .
\end{aligned}
\end{equation}

\bigskip
The (canonical) partition function of XXX${}_\frac12$ spin chain of length $L$ has been computed in \cite{Melzer:1993bi}. He also showed that the partition function gives the character of $\alg{su}(2)$ affine Kac-Moody algebra at level one in the large $L$ limit, as conjectured in \cite{Affleck:1985jc}. 
Their analysis slightly differs from ours in three points: we compute $\vev{M_2}$ rather than $\vev{e^{\beta M_2}}$, sum over the level-matched states, and consider the grand partition function by summing over $L$.\footnote{The level-matching condition may be included by introducing another chemical potential coupled to the total momentum.}

\section{Hagedorn transition}\label{sec:Hagedorn}

We compute the grand partition function of $\cN=4$ SYM in the $SU(2)$ sector at one-loop at large $N_c$ based on the above results. Then we determine the Hagedorn temperature of the $\cN=4$ SYM in the $SU(2)$ sector, namely the smallest temperature $T \ge 0$ at which the grand partition function diverges, as in \eqref{main TH}.
We will see that the Hagedorn temperature has numerous branches depending on the value of the chemical potential on the complex plane.

\subsection{Grand partition function}

Consider the grand partition function of $\cN=4$ SYM in \eqref{def:spt part fn}, 
\begin{equation}
\cZ (\beta, \vec \omega) = \tr \( e^{-\beta \fD + \sum_i \omega_i J_i} \), \qquad
\fD = \fD_0 + \lambda \fD_2 + \dots .
\end{equation}
The trace is taken over the Hilbert space of all gauge-invariant operators in the $SU(2)$ sector.
The partition function has the weak coupling expansion
\begin{equation}
\begin{aligned}
\cZ (\beta, \vec \omega) &= \tr \(
e^{-\beta \fD_0+ \sum_i  \omega_i J_i }
- \lambda \beta \, \fD_2 \, e^{-\beta \fD_0 + \sum_i \omega_i J_i }
+ \dots \)
\\
&= Z_0^{\rm MT} (x,y) - \frac{2 \, \lambda}{N_c} \, \beta \, Z_2^{\rm MT} (x,y)
+ \dots,
\end{aligned}
\label{spt part gen}
\end{equation}
where we used \eqref{def:mixing matrix}.
We assign the R-charge $J_i$ to complex scalars as
\begin{equation}
Z \, : \, J_i = \delta_{i1} \,, \quad
W \, : \, J_i = \delta_{i2} \,, \quad
N_Z = \frac{\fD_0 + J_1 - J_2}{2} \,, \quad
N_W = \frac{\fD_0 - J_1 + J_2}{2} .
\end{equation}
The partition function \eqref{def:spt part fn} depends on $(\beta, \omega_1, \omega_2)$, whereas the generating functions in \eqref{spt part gen} depends on $(x,y)$. The two sets of variables are related by\footnote{The tree-level grand partition \eqref{def:spt part fn} is invariant under the simultaneous shift of $(\beta,\omega_1,\omega_2)$ by $\nu$. This redundancy is broken at one-loop.}
\begin{equation}
x = e^{-\beta + \omega_1}, \qquad
y = e^{-\beta + \omega_2} , \qquad
w = e^{-\beta} \,, \qquad 
\beta = \frac{1}{T} \,.
\label{def:xyT}
\end{equation}
In particular, the low-temperature expansion corresponds to the expansion in small $x,y$.
The computation below is valid at large $N_c$\,, due to the non-perturbative corrections in \eqref{cZ finite Nc}.

\subsection{Hagedorn temperature}

We introduce
\begin{equation}
x \equiv e^{-\beta} \, \tilde x , \qquad
y \equiv e^{-\beta} \, \tilde y ,
\label{def:txy}
\end{equation}
and vary $T$ at a fixed $(\tilde x,\tilde y)$.
The tree-level generating function \eqref{Z0mt-2} has simple poles at
\begin{equation}
T_* = \frac{k}{\log \( \tilde x^k + \tilde y^k \)} \,,\qquad 
(k = 1, 2, \dots),
\label{location pole}
\end{equation}
and the one-loop generating function \eqref{Z2mt formula} has double poles at the same location. By using
\begin{equation}
\begin{gathered}
(\tilde x + \tilde y) - (\tilde x^k + \tilde y^k)^{1/k}
= \frac{ (\tilde x + \tilde y)^k - (\tilde x^k + \tilde y^k) }{\sum_{j=0}^{k-1} (\tilde x + \tilde y)^{k-1-j} (\tilde x^k + \tilde y^k)^{j/k}}
\ge 0,
\\[2mm]
{\rm for} \ \ (\tilde x, \tilde y) \in 
\cR_+ = \pare{ \tilde x \ge 0 \ \ {\rm and} \ \ \tilde y \ge 0, \ \tilde x + \tilde y \ge 1 }, 
\end{gathered}
\label{def:Rp}
\end{equation}
The term $k=1$ gives the smallest value of $T_*$ inside the region $\cR_+$\,. The condition $\tilde x + \tilde y \ge 1$ in $\cR_+$ guarantees $T_* \ge 0$.

We assume that the Hagedorn temperature and the partition function are expanded as
\begin{equation}
\begin{aligned}
T_H (\lambda) &= T_* \, \[ 1 + t_1 \, \lambda + O(\lambda^2) \] ,
\\[2mm]
\cZ (\beta, \vec \omega) &= \frac{c}{T - T_H (\lambda)} 
= \frac{c}{T-T_*} \[ 1 + \frac{\lambda T_* t_1}{T-T_*} + O(\lambda^2) \].
\end{aligned}
\end{equation}
Let us expand the partition function \eqref{spt part gen} around the pole \eqref{location pole} with $k=1$ and compare the result with the above expansion. We find that
\begin{equation}
T_H (\lambda) = \frac{1}{{\log (\tilde{x}+\tilde{y})}} \[ 1 + 
\frac{4 \lambda  \tilde{x} \tilde{y}}{\left(\tilde{x}+\tilde{y}\right)^2} \] ,\qquad
(\tilde x, \tilde y) \in \cR_+ \,.
\label{Hagedorn temp}
\end{equation}

Consider the region outside $\cR_+$.
When we cross the line $\tilde x + \tilde y = 1$, then $T_*$ becomes negative for all $k$. As we approach $\tilde x \to 0$ keeping $\tilde x + \tilde y \ge 1$, all simple poles in \eqref{location pole} accumulate at $T_* = 1/\log(\tilde y)$.
Let us take either $\tilde x$ and $\tilde y$ negative, where the chemical potentials \eqref{def:xyT} are shifted by $\pi i$.
When $|\tilde x|, |\tilde y|$ are large enough, the pole \eqref{location pole} with $k=2$ becomes the closest to the origin among those giving $T_* > 0$. Thus, in the region
\begin{equation}
\cR_- = \pare{ \tilde x \le 0 \ \ {\rm or} \ \ \tilde y \le 0, \ \tilde x^2 + \tilde y^2 \ge 1 }, 
\label{def:Rm}
\end{equation}
the one-loop Hagedorn temperature is given by
\begin{equation}
T_H (\lambda) = \frac{2}{\log (\tilde{x}^2+\tilde{y}^2 )} \[ 1 + 
\frac{4 \lambda  \tilde{x}^2 \tilde{y}^2}{(\tilde{x}^2+\tilde{y}^2)^2}
\] , \qquad 
(\tilde x, \tilde y) \in \cR_- \,.
\label{Hagedorn temp2}
\end{equation}
More generally, if we put $(\tilde x, \tilde y) \in \bb{C}^2$ on
\begin{equation}
{\rm Arg} \, \tilde x = \frac{2 \pi}{p_1} \,, \qquad
{\rm Arg} \, \tilde y = \frac{2 \pi}{p_2} \,, \qquad
p = {\rm lcm} (p_1, p_2), \qquad
(p_1, p_2 \in \bb{Z}_{\ge 1} \,, \ \ p \ll N_c^2),
\label{p-transition}
\end{equation}
the Hagedorn temperature is given by the pole \eqref{location pole} at $k = p$.
When $p = O(N_c^2)$, the Hagedorn transition may take place around $1/\log |\tilde x + \tilde y|$, because the free energy becomes $O(N_c^2)$ without hitting the pole.

\begin{figure}[t]
\begin{center}
\includegraphics[scale=0.77]{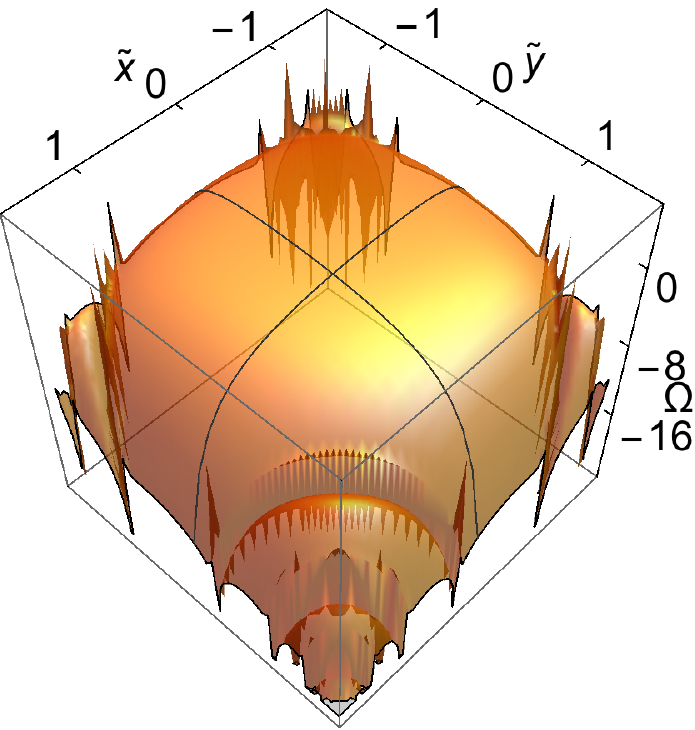}
\hspace{3mm}
\includegraphics[scale=0.7]{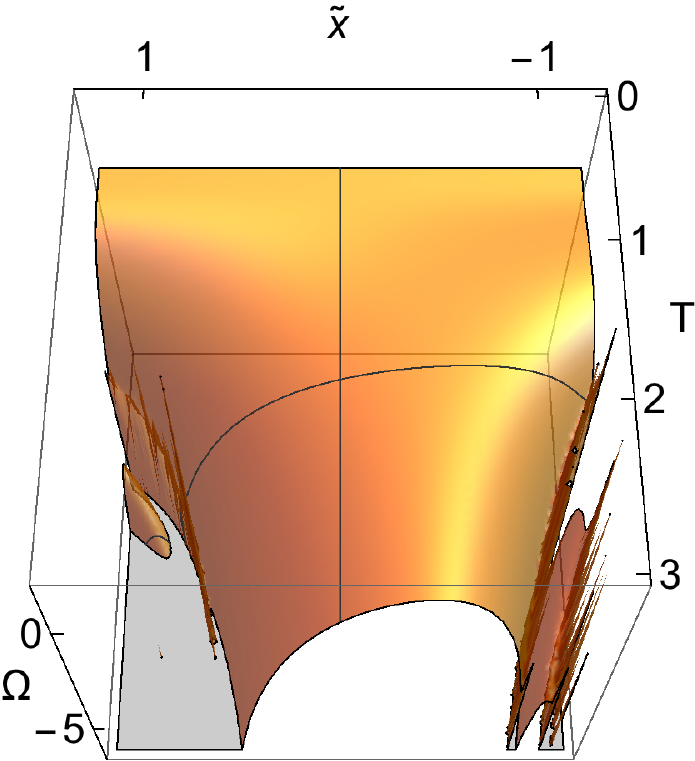}
\hspace{3mm}
\includegraphics[scale=0.7]{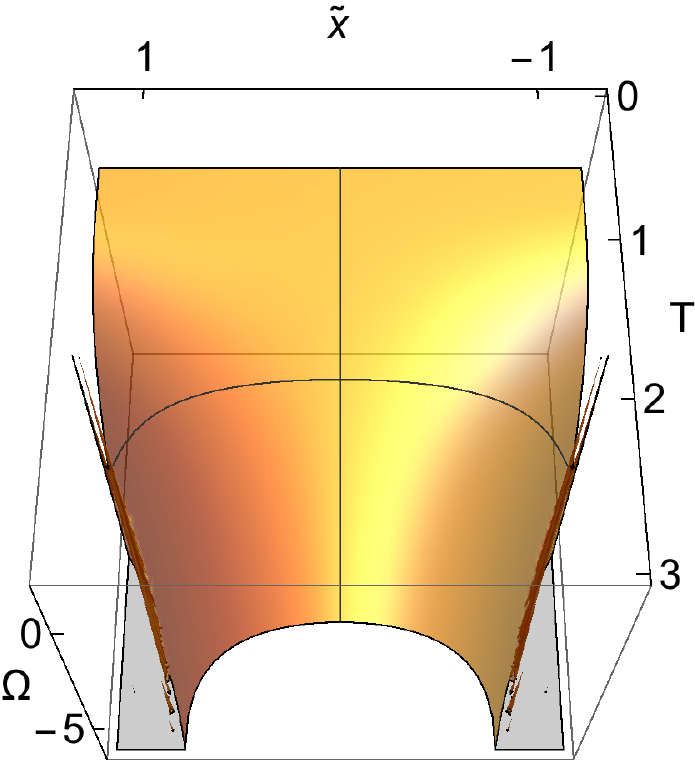}
\caption{Plots of ${\rm Re} \, \Omega$ at $\lambda=0.05$. The left figure shows the plot with $(T=5/3, \tilde x, \tilde y)$, the middle figure with $(T, \tilde x = \tilde y)$, and the right figure with $(T, \tilde x = - \tilde y)$.}
\label{fig:gnd part}
\end{center}
\end{figure}

In Figure \ref{fig:gnd part}, the plots of the grand partition function $\Omega \equiv - T \log \cZ$ are shown as a function of $(T, \tilde x, \tilde y)$. The left figure at a fixed $T$ shows that the singularity of $\Omega$ is associated with the boundary of $\cR_\pm$\,. By comparing the middle figure $(T, \tilde x = \tilde y)$ and the right figure $(T, \tilde x = - \tilde y)$, we find that the former is not invariant under the flip $\tilde x \leftrightarrow - \tilde x$, whereas the latter is invariant. This pattern is consistent with $\cR_\pm$\,.

\bigskip
For comparison with the literature, we vary $T$ at a fixed $(\tilde \omega_1 = \omega_1/\beta\,, \xi = y/x)$. It follows that
\begin{equation}
T_H (\lambda) = 
\begin{cases}
\ds \frac{1 - \tilde \omega_1}{\log (1+\xi)} \[ 1 + \frac{4 \lambda\, \xi}{(1+\xi)^2  \( 1 - \tilde \omega_1 \)} + \dots \] &\qquad (\tilde x, \tilde y) \in \cR_+ \,,
\\[5mm]
\ds \frac{2 \( 1 - \tilde \omega_1 \)}{\log (1+\xi^2)}  
\[ 1 + \frac{4 \lambda\, \xi ^2}{( 1+ \xi ^2)^2 \, (1- \tilde \omega_1)}
 + \dots \] &\qquad (\tilde x, \tilde y) \in \cR_- \,.
\end{cases}
\label{TH literature}
\end{equation}
The first line agrees with \cite{Spradlin:2004pp} when $\xi=1, \tilde \omega_1=0$, and with \cite{Harmark:2006di} when $\xi=1$.\footnote{Note that (1.2) of \cite{Spradlin:2004pp} is the Hagedorn temperature of the entire $\cN=4$ SYM. The Hagedorn temperature in the $SU(2)$ sector can be found e.g. in \cite{GomezReino:2005bq}.}

\bigskip
Let us make a few remarks on the Hagedorn transition.

First, the physical partition function should not diverge. This means that the system turns into the deconfined phase around the Hagedorn temperature, when the free energy becomes $O(N_c^2)$. 
In order to inspect the details of the phase transition, we need to evaluate \eqref{complete Z} in the large $N_c$ limit as in \cite{Sundborg:1999ue}. 
In the $SU(2)$ sector, it is expected that the system is described by $N_c^2+1$ harmonic oscillators around the Hagedorn temperature \cite{Harmark:2014mpa}.\footnote{The author thanks the referee of JHEP for this comment.}

Second, the parameter region $\tilde x<0$ or $\tilde y<0$ can be interpreted as the insertion of the number operator $(-1)^{N_Z}$ or $(-1)^{N_W}$ to the grand partition function \eqref{def:spt part fn}, which makes $Z$ or $W$ effectively a fermion. The pole at $k=1$ disappears when the scalar becomes fermionic. 
The pole at $k=2$ still contributes to the divergence because $Z^2$ or $W^2$ are bosonic. Similarly, when the transition takes place at $k=p$ as in \eqref{p-transition}, $Z^p, W^p$ are effectively bosonic.
This pattern indicates that only effective bosons form a condensate inside which $U(N_c)$ degrees of freedom are liberated from the confinement.

Third, the grand partition function at finite $N_c$ is a smooth function of the temperature, and no transition should happen \cite{Aharony:2003sx}. This can be checked by evaluating $\cZ^{SU(2)}_{N_c} (x,y)$ in \eqref{Molien-Weyl SUq}. 
For example, $\cZ^{SU(2)}_{N_c=2} (x,y)$ with $x=y=e^{-\beta}$ in \eqref{Nc=2 SU(2) exact} is regular for any $\beta > 0$. More generally, it is conjectured that the denominator of $\cZ^{SU(2)}_{N_c} (x,y)$ at any $N_c < \infty$ is always a product of the factors $(1-x^a y^b)$ for some integers $a,b \ge 0$ \cite{Djokovic}. Hence, the Hagedorn temperature is infinite at finite $N_c$\,.

\section{Conclusion and Outlook}

In this paper, we computed the grand partition function of $\cN=4$ SYM in the $SU(2)$ sector at one-loop by making use of finite group theory. Only the planar terms contribute in this setup, though our result is valid to all orders of perturbative $1/N_c$ expansion. We derived two expressions for the one-loop generating function, called Partition form and Totient form. Based on Totient form we computed the Hagedorn temperature which depends on general values of the chemical potentials. We argued how the Hagedorn temperature changes when the chemical potentials are complex.

As future directions of research, one can consider the grand partition functions in more general situations, such as finite $N_c$ corrections at one-loop, larger sectors of $\cN=4$ SYM, or higher order in $\lambda$ in the $SU(2)$ sector. It is also interesting to study superconformal field theories other than $\cN=4$ SYM, such as $\beta$-deformed and $\gamma$-deformed theories \cite{Fokken:2014moa}, ABJM model \cite{Smedback:2010ji}, theories with 16 supercharges \cite{Grignani:2007xz}, and quiver gauge theories \cite{Pasukonis:2013ts}. Our counting methods should be applicable to integrable models like $q$-deformed Hubbard model \cite{Beisert:2008tw}, which is a generalization of XXZ spin chain.

Another topic is to develop group-theoretical techniques to study multi-point functions. It is well known that the OPE limit of four-point functions in $\cN=4$ SYM yields the sum of the anomalous dimensions of intermediate operators, weighted by the square of OPE coefficients. Such objects have been studied by conformal bootstrap \cite{Dolan:2004iy} and integrability methods \cite{Eden:2016xvg,Fleury:2016ykk,Basso:2017khq}. 
The effects of $1/N_c$ corrections in such a limit is worth further investigation.

\subsubsection*{Acknowledgements}

This work is supported by FAPESP grants 2011/11973-4, 2015/04030-7 and 2016/01343-7, 2016/25619-1.
RS thanks Sanjaye Ramgoolam for discussions and comments on the manuscript.

\appendix

\section{Notation}\label{app:notation}

\begin{figure}[t]
\begin{center}
\includegraphics[scale=1.2]{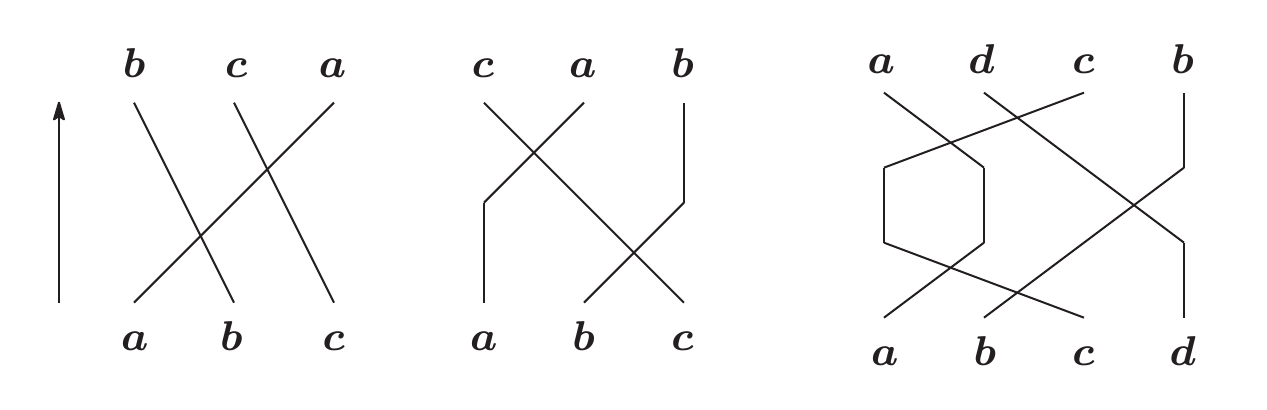}
\caption{Permutations represented as diagrams. Left Figure shows $(123) \in S_3$\,, Middle Figure shows $(12)(23)=(132)$, and Right Figure shows $(123)(34)(132)=(24)$.}
\label{fig:notation}
\end{center}
\end{figure}

A permutation cycle is denoted by $(i_1 i_2 \dots i_\ell)$. We define the action of permutation by keeping track of the position of a list. Algebraically, it means
\begin{equation}
\sigma \,:\, \{ v (1), \dots, v (n) \} \ \mapsto \ \{v' (1), \dots , v' (n) \}, \qquad
v'(n) = v(\sigma(n)) .
\end{equation}
For example, we have
\begin{equation}
\sigma = (123) \,:\, v = \{a,b,c\} \ \mapsto \ v' = \{b,c,a\},
\end{equation}
as shown in Figure \ref{fig:notation}. As corollaries, we find
\begin{equation}
\begin{gathered}
(12) \cdot (23) = (132), \qquad
(132) \cdot \{ 1 \} = (23) \cdot \Big( (12) \cdot \{ 1 \} \Big) = (23) \cdot \{2\} 
= \{ 3 \},
\\
(123) \cdot (34) \cdot (132) = (24),
\end{gathered}
\end{equation}
and in general
\begin{equation}
\alpha \cdot \beta \( n \) = \beta ( \alpha (n) ), \qquad
\alpha^{-1} (ij) \, \alpha = (\alpha(i) \alpha(j)).
\label{perm identities}
\end{equation}

In {\tt Mathematica}, {\tt Permute[]} gives $v' (n) = v (\sigma(n))$. For example,
\begin{gather}
\text{\tt Permute[\{a,b,c\},Cycles[\{\{1,2,3\}\}]] = \{c,a,b\}} 
\notag \\[1mm]
\text{\tt PermutationReplace[\{1,2,4\},Cycles[\{\{1,2,3\}\}]] = \{2,3,4\}} 
\\[1mm]
\text{\tt PermutationProduct[Cycles[\{\{3,4\}\}],Cycles[\{\{1,2,3\}\}]] = Cycles[\{\{1,2,3,4\}\}]} 
\notag
\end{gather}

\section{Details of derivation}

The second term of \eqref{su2 1loop av mn} can be rewritten as
\begin{equation}
\vev{M_2}_{m,n}^{\rm (2nd)} 
= \frac{N_c}{m! \, n!} \, \sum_{\alpha} 
\sum_{i \neq j=1}^{L} \sum_{\mu \in S_m \times S_n} 
\delta(i \alpha(j)) \, 
\delta_{L} \Big( \mu_0 \alpha^{-1} \mu_0^{-1} \alpha \, \Big) , \qquad
\mu_0 = (ij) \, \mu .
\label{One-loop mixing 2nd}
\end{equation}
This quantity will be computed below. 
The result is called Partition form when we sum over $\alpha \in S_L$, and Totient form when $\alpha \in \bb{Z}_L$\,.

\subsection{Second term in Partition form}\label{app:part 2nd}

We evaluate $\vev{M_2}_{m,n}^{\rm (2nd)}$ in the following steps:
\begin{center}
\fboxsep = 6pt
\fbox{\parbox{0.85 \textwidth}{
\addtolength{\leftskip}{7mm}
1) Choose $\mu \in T_p \times T_q \subset S_m \times S_n$

2) Generate $\mu_0 = (ij) \, \mu$ by summing over $(i,j)$ 

3) Solve the two $\delta$-function constraints simultaneously
}}
\end{center}

\medskip \noindent
1) We denote the cycle type of $\mu$ by $r \vdash L$. We have $r_k = p_k + q_k$ for $1 \le k \le L$, and
\begin{equation}
\sum_{\mu \in S_m \times S_n} f(\mu) =
\sum_{p \vdash m} \sum_{q \vdash n} |T_p| |T_q| f( [\mu] ) .
\end{equation}

\medskip \noindent
2) The cycle type of $\mu_0$ depends on how $i$ and $j$ appear inside $\mu$.
Let us parametrize the permutation cycles of $\mu$ which contain $i \,, j$ by $(x_1 \dots x_a)$. We write $(y_1 \dots y_b)$ if $i,j$ belong to different cycles, including the case $a=b$. Then
\begin{equation}
(i j) \, \mu\sim \begin{cases}
(i j) \, (x_1 \dots x_{l-1} \, i \, x_{l+1} \dots x_{a-1} j) 
= (x_1 \dots x_{l-1} \, i) (x_{l+1} \dots x_{a-1} \, j) 
\\[2mm]
(i j) \, (x_1 \dots x_{a-1} i) (y_1 \dots y_{b-1} j) 
= (x_1 \dots x_{a-1} \, i \, y_1 \dots y_{b-1} \, j)
\end{cases}
\label{mu to mu0}
\end{equation}
Thus, the transposition $(ij)$ relates the cycle type of $\mu$ and $\mu_0$ as
\begin{alignat}{9}
\{ r_l \,, r_{a-l} \,, r_a \} &\to \{ r_l + 1 , r_{a-l} +1, r_a -1 \}, &\qquad
{\rm if} \ \ j &= \mu^{a-l} (i) &\qquad &(1 \le l \le a-1), 
\label{mu to mu0-a} \\[1mm]
\{ r_a \,, r_b \,, r_{a+b} \} &\to \{ r_a -1 , r_b -1 , r_{a+b} +1 \}, &\qquad
{\rm if} \ \ j &\neq \mu^m (i) &\qquad &(\forall m \in \bb{Z}),
\label{mu to mu0-b} \\[1mm]
\{ r_a \,, r_{2a} \} &\to \{ r_a - 2 , r_{2a} +1 \}, &\qquad
{\rm if} \ \ j &\neq \mu^m (i) &\qquad &(\forall m \in \bb{Z}).
\label{mu to mu0-c}
\end{alignat}

Let us count the number of $(i,j)$ corresponding to each line of \eqref{mu to mu0}.
As for the first line, i.e. splitting, we choose a cycle of length $a$ and split it into $l + (a-l)$, for $1 \le l \le a-1$. There are $a r_a$ ways to choose $i$, and the choice of $j$ is unique for a given $(i, l)$. 
As for the  second line, i.e. joining, we choose two different cycles of length $a$ and $b$.  If $a \neq b$, there are $a b \, r_a r_b$ ways to identify $i \,, j$\,.
And if $a=b$, there are $a^2 \, r_a (r_a-1)$ ways. 
In total, we have\footnote{We used $\sum_a r_a = L$ and $r_c=0$ for $c > L$.}
\begin{equation}
\begin{aligned}
&\sum_{a=1}^{L} \sum_{l=1}^{a-1} a \, r_a
+ \sum_{a \neq b = 1}^{L} a b \, r_a r_b
+ \sum_{a=1}^{L} a^2 \, r_a (r_a-1) ,
\\
&= \sum_{a=1}^{L} a (a-1) \, r_a
+ \sum_{a, b = 1}^{L} a b \, r_a r_b
- \sum_{a=1}^{L} a^2 \, r_a 
\\
&= L(L-1).
\end{aligned}
\end{equation}
Thus, we replace the sum over $(i,j)$ in \eqref{One-loop mixing 2nd} by the sums over $a, b, l$ shown above. Schematically, the sum of dimensions is given by
\begin{equation}
\vev{M_2}_{m,n}^{\rm (2nd)} 
=
\[ \, r_a \to r_l + r_{a-l} \, \]_{\rm split}
+
\[ \, r_a + r_b \to r_{a+b} \, \]_{\rm join} 
+
\[ \, 2 \, r_a \to r_{2a} \, \]_{\rm join} \,.
\end{equation}

\medskip \noindent
3) We label all possible choices of $(i,j)$ by $\zeta = 1,2, \dots , L(L-1)$, and denote the cycle type of $\mu_0^{(\zeta)}$ by $\bar r^{(\zeta)}$.
For each $\bar r^{(\zeta)}$ we choose $\alpha$ as
\begin{equation}
\alpha \in {\rm Stab} (\mu_0^{(\zeta)}) = \prod_{k=1}^{L} S_{\bar r^{(\zeta)}_k} [ \bb{Z}_k ]  ,
\end{equation}
to solve the $\delta$-function constraint \eqref{One-loop mixing 2nd}. The $\zeta$ belongs to either of the two groups in \eqref{mu to mu0}.

Next, we solve the planarity condition $i = \alpha (j)$ for the three cases \eqref{mu to mu0-a}-\eqref{mu to mu0-c}. Recall that $i$ and $j$ belong to the cycle of the same length in $\mu_0$ to solve the conditions $i = \alpha (j)$ and $\alpha \in {\rm Stab} (\mu_0)$, as discussed in Section \ref{sec:part 1st}.

As for the splitting case \eqref{mu to mu0-a}, only the process $r_{2l} \to r_l + r_l$ can solve the condition $i = \alpha (j)$. The stabilizer of $\mu_0$ is 
\begin{equation}
{\rm Stab} (\mu_0)_{\rm split} = S_{r_l+2} [\bb{Z}_l] \cdot S_{r_{2l}-1} [ \bb{Z}_{2l} ] \cdot \prod_{k \neq l, 2l} S_{r_k} [ \bb{Z}_k ] .
\label{stab split}
\end{equation}
There are $2l \, r_{2l}$ choices of $i,j$, including the interchange $i \leftrightarrow j$.\footnote{The interchange $i \leftrightarrow j$ does not change the transposition $(ij)$, but we need to sum over all $i \neq j$ in \eqref{One-loop mixing 2nd}.}
In order to solve $i=\alpha(j)$ when $i,j$ appear in the cycles of length $l$, we write
\begin{equation}
\alpha = (ij) \alpha_0 \,, \qquad \alpha_0 (i) = i .
\end{equation}
Here $\alpha_0$ freezes the cycle of $i$ but not of $j$, which restricts ${\rm Stab} (\mu_0)_{\rm split}$ down to
\begin{equation}
\alpha_0 \in {\rm Stab}' (\mu_0)_{\rm split} = S_{r_l+1} [\bb{Z}_l] \cdot S_{r_{2l}-1} [ \bb{Z}_{2l} ] \cdot \prod_{k \neq l, 2l} S_{r_k} [ \bb{Z}_k ] .
\label{stab' split}
\end{equation}
The number of solutions in the splitting process is given by
\begin{equation}
N'_{sol} (\mu)_{\rm split} = \sum_{l=1}^{L/2} \theta_> (r_{2l}) \, 2l \, r_{2l} \,
\Big| {\rm Stab}' (\mu_0)_{\rm split} \Big|
= \sum_{l=1}^{L/2} \theta_> (r_{2l}) \, l \, (r_l+1) \Big| {\rm Stab} (\mu) \Big| .
\label{N'split}
\end{equation}

As for the joining cases \eqref{mu to mu0-b} and \eqref{mu to mu0-c}, any $a,b$ can solve the condition $i=\alpha(j)$. The stabilizer of $\mu_0$ is 
\begin{equation}
{\rm Stab} (\mu_0)_{\rm join} = \begin{cases}
\ds S_{r_a-1} [\bb{Z}_a] \cdot S_{r_b-1} [\bb{Z}_b] \cdot S_{r_{a+b}+1} [\bb{Z}_{a+b}] \prod_{k \neq a,b,a+b} S_{r_k} [ \bb{Z}_k ]
&\quad (a \neq b) 
\\[3mm]
\ds S_{r_a-2} [\bb{Z}_a] \cdot S_{r_{2a}+1} [ \bb{Z}_{2a} ] \cdot \prod_{k \neq a,2a} S_{r_k} [ \bb{Z}_k ] 
&\quad (a=b) .
\end{cases} 
\label{stab join}
\end{equation}
For $a \neq b$ there are $a b \, r_a r_b$ choices of $i,j$, and for $a=b$ there are $a^2 \, r_a (r_a-1)$ choices, including the interchange $i \leftrightarrow j$.
In order to solve $i=\alpha(j)$ when $i,j$ appear in the cycle of length $a+b$, we restrict ${\rm Stab} (\mu_0)_{\rm join}$ down to
\begin{equation}
\alpha_0 \in {\rm Stab}' (\mu_0)_{\rm join} = \begin{cases}
\ds S_{r_a-1} [\bb{Z}_a] \cdot S_{r_b-1} [\bb{Z}_b] \cdot \prod_{k \neq a,b} S_{r_k} [ \bb{Z}_k ]
&\quad (a \neq b) 
\\[2mm]
\ds S_{r_a-2} [\bb{Z}_a] \cdot \prod_{k \neq a} S_{r_k} [ \bb{Z}_k ] 
&\quad (a=b).
\end{cases}
\label{stab' join}
\end{equation}
The number of solutions in the joining process is given by
\begin{align}
N'_{sol} (\mu)_{\rm join} &= \sum_{\substack{a \neq b \\ a+b \le L}} \theta_> (r_a) \theta_> (r_b) \, a b \, r_a r_b \, \Big| {\rm Stab}' (\mu_0)_{\rm join}^{a \neq b} \Big|
\notag \\
&\quad 
+ \sum_{a=1}^{L/2} \theta_> (r_a-1) \, a^2 r_a (r_a-1) \, \Big| {\rm Stab}' (\mu_0)_{\rm join}^{a = b} \Big|
\notag \\
&= \Bigl\{ \sum_{\substack{a \neq b \\ a+b \le L}} \theta_> (r_a) \theta_> (r_b) 
+ \sum_{a=1}^{L/2} \theta_> (r_a-1) 
\Bigr\} \, \Big| {\rm Stab} (\mu) \Big|  .
\label{N'join}
\end{align}

\medskip \noindent
4) In total, we have
\begin{equation}
\begin{aligned}
N'_{sol} &\equiv N'_{sol} (\mu)_{\rm split} + N'_{sol} (\mu)_{\rm join}
= \Theta (r) \, \Big| \prod_{k} S_{r_k} [ \bb{Z}_k ] \Big|,
\\
\Theta (r) &\equiv 
\sum_{a=1}^{L/2} a \, (r_a+1) \theta_> (r_{2a}) 
+ \sum_{a< b}^L 2 \, \theta_> (L+1-a-b) \theta_> (r_a) \theta_> (r_b) 
+ \sum_{a=1}^{L/2} \theta_> (r_a-1) ,
\end{aligned}
\end{equation}
where we removed the constraint $a+b \le L$ by inserting $\theta_> (L+1-a-b)$.
Following \eqref{orbit stabilizer}-\eqref{Z0mt-2}, we obtain the generating function for the second term as
\begin{equation}
\begin{aligned}
Z_2^{SU(2), {\rm (2nd)}} (x, y) &= N_c \, \sum_{m,n} \sum_{p \vdash m, q \vdash n} |T_p| |T_q| \, \frac{x^m y^n}{m! \, n!} \, N_{sol} \,,
\\
&= N_c \, \sum_{L=0}^\infty \sum_{r \vdash L} \prod_{k=1} (x^k + y^k)^{r_k} \, \Theta (r) .
\end{aligned}
\label{gen av 2nd}
\end{equation}

\bigskip
Consider an example. Let $\mu$ be $(1)(23)(456)$, having $r=[1^1,2^1,3^1] \vdash 6$. Then ${\rm Stab} (\mu) = \bb{Z}_1 \cdot \bb{Z}_2 \cdot \bb{Z}_3$\,, which has the order $1 \times 2 \times 3 = 6$. The possible splitting process is
\begin{equation}
\mu_0 = (23) \mu = (32) \mu = (1)(2)(3)(456), \qquad
{\rm Stab} (\mu_0) = S_3 [\bb{Z}_1] \cdot \bb{Z}_3 \,.
\end{equation}
We have $(i,j) = (2,3)$ or $(3,2)$. The list of $\{ \alpha(j) \, | \, \alpha \in {\rm Stab} (\mu_0) \}$ is $\{ 1^6, 2^6, 3^6 \}$ for $j=2,3$, in the notation of Table \ref{tab:alpha j}. Thus, the number of solutions is
\begin{equation}
N'_{sol} (\mu)_{\rm split} = 6 \times 2 = 2 \, \Big| {\rm Stab} (\mu) \Big|,
\end{equation}
which agrees with \eqref{N'split}.
The possible joining processes are
\begin{equation}
\mu_0 = \begin{cases}
(123)(456), \dots & \qquad (ij) = (12), (13) \\
(23)(1456), \dots & \qquad (ij) = (14), (15), (16) \\
(1)(23456), \dots &\qquad (ij) = (24), (25), (26), (34), (35), (36),
\end{cases}
\end{equation}
and their stabilizers are
\begin{equation}
{\rm Stab} (\mu_0) = \begin{cases}
S_2 [\bb{Z}_3] &\qquad \bar r = [3^2] \\
\bb{Z}_2 \cdot \bb{Z}_4 &\qquad \bar r = [2^1,4^1] \\
\bb{Z}_1 \cdot \bb{Z}_5 &\qquad \bar r = [1^1, 5^1].
\end{cases}
\end{equation}
The lists of $\alpha(j)$ is
\begin{alignat}{9}
j &= 2, &\qquad \alpha(j) &= \pare{ 1^3, 2^3, 3^3, 4^3, 5^3, 6^3 }, &\qquad \bar r &= [3^2] \\
j &= 4, &\qquad \alpha(j) &= \pare{ 1^2, 4^2, 5^2, 6^2 }, &\qquad \bar r &= [2^1,4^1] \\
j &= 4, &\qquad \alpha(j) &= \pare{ 2^1, 3^1, 4^1, 5^1, 6^1 }, &\qquad \bar r &= [1^1, 5^1].
\end{alignat}
Thus, the number of solutions is
\begin{equation}
N'_{sol} (\mu)_{\rm join} = 3 \times 4 + 2 \times 6 + 1 \times 12 = 36
= 6 \, \Big| {\rm Stab} (\mu) \Big|,
\end{equation}
which agrees with \eqref{N'join}.

\subsection{Second term in Totient form}\label{app:totient 2nd}

We evaluate $\vev{M_2}_{m,n}^{\rm (2nd)}$ in the following steps:
\begin{center}
\fboxsep = 6pt
\fbox{\parbox{0.85 \textwidth}{
\addtolength{\leftskip}{7mm}
1) Choose $\mu \in T_p \times T_q \subset S_m \times S_n$

2) Generate $\mu = (ij) \, \mu_0$ from $\mu_0$ in $\bb{Z}_d^\ell$

3) Generate $\mu_0 = (ij) \, \mu$ by reverting the last step

4) Solve the two $\delta$-function constraints simultaneously
}}
\end{center}

\medskip \noindent
1) We denote the cycle type of $\mu \in S_m \times S_n$ by $p \vdash m$ and $q \vdash n$. Let $m,n$ be divisible by $d \ge 1$ as in \eqref{def:dmn}.

\medskip \noindent
2) Given $\mu_0 \in \bb{Z}_d^\ell$ parametrized as in \eqref{parameter tilde mu}, we generate $\mu = (ij) \, \mu_0$\,. We classify two cases depending on whether $i,j$ belong to the same or different cycles of $\mu_0$\,,
\begin{equation}
\begin{aligned}
{\rm Same:} \qquad
(i,j) &= (\tilde m_{kh}, \tilde m_{kd}), \qquad (1 \le k \le \ell, \ 2 \le h \le d)
\\
{\rm Different:} \qquad
(i,j) &= (\tilde m_{kd}, \tilde m_{k'd}), \qquad (1 \le k \neq k' \le \ell)
\end{aligned}
\label{choose ij}
\end{equation}
which correspond to
\begin{align}
(ij) (\tilde m_{k1} \dots \tilde m_{kd}) &\to (\tilde m_{k1} \dots \tilde m_{k,h-1} \, i) (\tilde m_{k, h+1} \dots \tilde m_{k,d-1} \, j), \qquad
\label{choose ij Same} \\[1mm]
(ij) (\tilde m_{k1} \dots \tilde m_{kd}) (\tilde m_{k'1} \dots \tilde m_{k'd}) &\to
(\tilde m_{k1} \dots \tilde m_{kd} \, i \, \tilde m_{k'1} \dots \tilde m_{k' d-1} \, j).
\label{choose ij Different} 
\end{align}
In terms of cycle types of $\mu_0$ and $\mu$, these processes can be written as
\begin{equation}
\begin{aligned}
\text{Same:} &\quad [d^\ell] \ \to \ [h,d-h,d^{\ell-1}]
\\
\text{Different:} &\quad [d^\ell] \ \to \ [d^{\ell-2}, 2d]
\end{aligned}
\label{choose ij-2}
\end{equation}
We assume $\ell=1$ in the Same case and $\ell \ge 2$ in the Different case, because we will see later that other cases do not contribute. These conditions are equivalent to $d=L$ and $d<L$, respectively.

\medskip \noindent
3) We revert the above argument, and generate $\mu_0 \in \bb{Z}_d^\ell$ from $\mu \in S_m \times S_n$\,. For the Same case, we choose
\begin{equation}
\mu \in \begin{cases}
\bb{Z}_m \times \bb{Z}_n \,, &\quad (mn \neq 0) \\
\bb{Z}_h \times \bb{Z}_{L-h} &\quad (mn = 0),
\end{cases}
\qquad
\mu_0 \in \bb{Z}_{L} \,.
\label{split mu to mu0}
\end{equation}
First, consider the case $mn \neq 0$.
There are $(m-1)! (n-1)!$ ways to choose $\bb{Z}_m \times \bb{Z}_n$ from $S_m \times S_n$\,. Then we choose $(i,j)$ from $(\bb{Z}_m \,, \bb{Z}_n)$ or $(\bb{Z}_n \,, \bb{Z}_m)$. Using the formula
\begin{equation}
(m, L) (1, \cdots , m) (m+1, \cdots, L) = (1, \cdots , m, m+1, \cdots, L) \in \bb{Z}_{L} \,,
\end{equation}
we generate $(ij) \mu = \mu_0 \in \bb{Z}_{L}$\,. There are $2mn$ ways to choose $(i,j)$, giving us the multiplicity
\begin{equation}
(m-1)! (n-1)! \cdot 2mn = 2 m! n!.
\label{count Same-1}
\end{equation}
For any choices we obtain 
\begin{equation}
j = \begin{cases}
\mu_0^n (i) &\qquad \({\rm if} \ \ m+1 \le j \le L\) \\
\mu_0^m (i) &\qquad \({\rm if} \ \ 1 \le j \le n \).
\end{cases}
\label{split ij rel}
\end{equation}
Next, consider the case $mn=0$. The number of choices of $\mu \in \bb{Z}_h \times \bb{Z}_{L-h}$ from $S_{L}$ is
\begin{equation}
\frac{L!}{h \, (L-h)} \quad \(1 \le h < \frac{L}{2} \), \qquad
\frac{L!}{2 (L/2)^2} \quad \(h= \frac{L}{2} \),
\label{choices ZhZL}
\end{equation}
and there are $2 h (L-h)$ ways to choose $(i,j)$ for any $h$, giving the multiplicity
\footnote{We added an extra factor of $1/2$ by extending the summation range of $h$.}
\begin{equation}
\frac{L!}{2 h \, (L-h)} \cdot 2 h (L-h) = L!, \qquad
\( h=1,2, \dots L-1 \).
\label{count Same-2}
\end{equation}
For any choices, we have $j = \mu_0^h (i)$ or $\mu_0^{L-h} (i)$.

For the Different case, we set
\begin{equation}
\mu \in \( \bb{Z}_d^{m'-2} \times \bb{Z}_{2d} \) \times \bb{Z}_d^{n'} \quad {\rm or} \quad
\bb{Z}_d^{m'} \times \( \bb{Z}_d^{n'-2} \times \bb{Z}_{2d} \), \qquad
\mu_0 \in \bb{Z}_d^{m'+n'} = \bb{Z}_d^\ell \,.
\end{equation}
The case with $mn=0$ is allowed.
The number of choices of $\( \bb{Z}_d^{m'-2} \times \bb{Z}_{2d} \) \times \bb{Z}_d^{n'}$ from $S_m \times S_n$ is
\begin{equation}
\frac{|S_m \times S_n|}{|{\rm Stab} ( \bb{Z}_d^{m'-2} \times \bb{Z}_{2d} ) \times {\rm Stab} ( \bb{Z}_d^{n'} )|} = 
\frac{m! n!}{d^{m'+n'-2} \, (2d) \, (m'-2)! \, (n')!} \,.
\end{equation}
The other case $m' \leftrightarrow n'$ can be treated similarly.
Then we choose $(i,j)$ from $\bb{Z}_{2d}$ and use the formula
\begin{equation}
(m_d \, m_{2d}) (m_1 \, m_2 \dots m_{2d}) = (m_1 \, m_2 \dots m_d) (m_{d+1} \, m_{d+2} \dots m_{2d}) \in \bb{Z}_d^2 \,,
\end{equation}
to generate $\mu_0 \in \bb{Z}_d^\ell$\,. There are $2d$ ways to choose such $(i,j)$ from $\bb{Z}_{2d}$\,, which gives the multiplicity
\begin{equation}
\frac{m! n!}{d^{m'+n'-2} \, (2d) \, (m'-2)! \, (n')!} \, (2d)
= \frac{m! n!}{d^{\ell-2} \, (m'-2)! \, (n')!}
\end{equation}

\medskip \noindent
4) Given $(i,j,\mu_0)$, we construct $\alpha$ via \eqref{sol:stabilizer st}. This equation is repeated below:
\begin{equation}
\begin{gathered}
\alpha = \Bigl( a_1 \dots a_\ell \ \tilde \mu^\kappa (a_1) \dots \tilde \mu^\kappa (a_\ell) \, \tilde \mu^{2\kappa} (a_1) \dots \tilde \mu^{2\kappa} (a_\ell) \dots
\tilde \mu^{(d-1)\kappa}(a_1) \dots \tilde \mu^{(d-1)\kappa} (a_\ell) \Bigr) ,
\\
1 \le \kappa < d, \qquad {\rm gcd} (\kappa, d) =1.
\end{gathered}
\label{stabilizer st2}
\end{equation}
Now we solve $i = \alpha(j)$, keeping in mind that $(a_1, a_2, \dots a_\ell)$ in \eqref{stabilizer st2} belong to the different cycles of $\tilde \mu \equiv \mu_0$\,.

For the Same case, namely when $i,j$ belong to the same cycle of $\mu_0$\,, the equation $i=\alpha(j)$ has a solution only if
\begin{equation}
\ell =1, \qquad
d = L, \qquad
i  = \tilde \mu^\kappa (j) \qquad \Leftrightarrow \qquad 
\alpha = \tilde \mu^\kappa.
\end{equation}
From \eqref{split ij rel} we find $\kappa = m$ or $n$ for $mn \neq 0$, and $\kappa = h, L-h$ for $mn=0$.

For the Different case, the general solution is
\begin{equation}
\begin{gathered}
j = \tilde \mu^{\omega \kappa} (a_\xi ), \quad
i = \tilde \mu^{\omega \kappa} (a_{\xi+1}), \qquad {\rm or} \qquad
j = \tilde \mu^{\omega \kappa} (a_\ell ), \quad
i = \tilde \mu^{(\omega + 1) \kappa} (a_1),
\\[1mm]
\ell \ge 2, \quad d \le L/2,\quad
(1 \le \xi \le \ell-1, \ 0 \le \omega \le d-1) .
\end{gathered}
\label{gen ij diff}
\end{equation}
We use the overall translation of $\alpha \in \bb{Z}_L$ to fix $j=a_1$ and $i=a_2$\,, for both solutions in \eqref{gen ij diff}. Then $\alpha$ is given by
\begin{equation}
\alpha = 
(j \, i \, a_3 \dots a_\ell \, \tilde \mu(j) \tilde \mu^\kappa (i) \dots \tilde \mu^\kappa (a_\ell) \dots \tilde \mu^{(d-1)\kappa} (j) \dots \tilde \mu^{(d-1)\kappa} (a_\ell))
\end{equation}
The number of choices of $a_3 \dots a_\ell$ or $a_2 \dots a_{\ell-1}$ is
\begin{equation}
d^{\ell-2} \, (\ell - 2)! 
\end{equation}

\medskip \noindent
5) We summarize the above calculation. The second term of the sum of dimensions consists of two parts:
\begin{equation}
\vev{M_2}_{m,n}^{\rm (2nd)} 
= \frac{N_c}{m! n!} \( N_{m,n}^{\rm Same} + N_{m,n}^{\rm Different} \).
\label{gen 1L-ST 2nd}
\end{equation}
The number of solutions in the splitting case is
\begin{equation}
N_{m,n}^{\rm Same} = 
\begin{cases}
2 \, m! n! \, \delta({\rm gcd}(m,n), 1) , &\quad (mn \neq 0)
\\
\ds L! \, {\rm Tot} (L)  &\quad (mn = 0).
\end{cases}
\label{Nmn split st}
\end{equation}
where we used \eqref{def:Totient} in the last line.
The number of solutions in the joining case is\footnote{If $m' \le 1$ or $n' \le 1$, one of the two terms in \eqref{Nmn join st} vanishes due to $(-1)!=\infty$.}
\begin{equation}
\begin{aligned}
N_{m,n}^{\rm Different} &= \sum_{\substack{d=1 \\ d|m, \ d|n}}^{L-1} 
\frac{m! n!}{d^{\ell-2} \, (m'-2)! \, (n')!} \, d^{\ell-2} \, (\ell - 2)! \, {\rm Tot} (d) + \( m' \leftrightarrow n' \) ,
\\
&= \sum_{\substack{d=1 \\ d|m, \ d|n}}^{L-1} m! n! \, {\rm Tot} (d) \pare{
\frac{(L/d-2)!}{(m/d-2)! (n/d)!} + \frac{(L/d-2)!}{(m/d)! (n/d-2)!} } .
\end{aligned}
\label{Nmn join st}
\end{equation}

\bigskip
Consider some examples. Suppose $(m,n)=(7,2)$ and consider the Same case, 
\begin{equation}
\mu \in \bb{Z}_7 \times \bb{Z}_2 \,, \qquad
\mu_0 \in \bb{Z}_9 \,.
\end{equation}
There are $6! \times 1! = 720$ ways to choose such $\mu$ from $S_7 \times S_2$\,. Then we choose $(i,j)$ from two cycles, like
\begin{equation}
(ij) \mu = (79) (1234567)(89) = (123456789)  = \mu_0 \in \bb{Z}_9 \,.
\end{equation}
There are $2 \times 7 \times 2 = 28$ ways to choose $(i,j)$, and all of them satisfy $j = \mu_0^2 (i)$ or $j = \mu_0^7 (i)$.
The solution of $\mu_0 \alpha^{-1} \mu_0^{-1} \alpha = 1$ is
\begin{equation}
\alpha = (a \, \mu_0^\kappa (a) \, \mu_0^{2\kappa} (a) \, \dots \mu_0^{8 \kappa} (a) ),
\qquad
{\rm gcd} (\kappa, 9) = 1.
\end{equation}
The condition $i = \alpha(j)$ requires $\kappa = 2, 7$. Therefore, the number of Same solutions is
\begin{equation}
720 \times 28 = 20160 = 7! \times 2! \times 2,
\end{equation}
in agreement with \eqref{Nmn split st}.

If $\mu \in S_6 \times S_3$\,, then there is no contribution from the Same case due to ${\rm gcd} (3,6) \neq 1$.
In the Different case we can generate $\mu_0 \in \bb{Z}_3^3$\,.
There are $5! \times 2! = 240$ ways to choose $\mu \in \bb{Z}_6 \times \bb{Z}_3$ from $S_6 \times S_3$. 
We choose $(i,j)$ from $\bb{Z}_6$\,, like
\begin{equation}
(ij) \mu = (36) (123456)(789) = (123)(456)(789) = \mu_0 \in \bb{Z}_3^3 \,.
\end{equation}
There are 6 choices of $(i,j)$.
The solution of $\mu_0 \alpha^{-1} \mu_0^{-1} \alpha = 1$ can be written as
\begin{equation}
\alpha = (a_1 a_2 a_3 \, \mu_0^\kappa (a_1) \, \mu_0^\kappa (a_2) \, \mu_0^\kappa (a_3)
\dots \mu_0^{2 \kappa} (a_3) ),
\qquad
\kappa \in \{1,2\}.
\end{equation}
We look for the solutions of $i = \alpha(j)$. By overall translation we put 
\begin{equation}
(i,j) = (a_2, a_1).
\end{equation}
If $(i,j)=(3,6)$, there are 3 ways to choose $a_3$ from $(789)$. The multiplicity for other $(i,j)$ is identical.
Therefore, the number of Different solutions is
\begin{equation}
240 \times 6 \times 3 \times {\rm Tot} (3) = 6! \times 3! \times 2,
\end{equation}
which agrees with \eqref{Nmn join st}.

\bibliographystyle{utphys}
\bibliography{bibmix}{}

\end{document}